\documentclass{article}

\usepackage{siunitx}
\usepackage[utf8]{inputenc}
\usepackage{cite}
\usepackage{amsmath}
\usepackage{amssymb}
\usepackage{graphicx}
\usepackage{subcaption}
\usepackage{dsfont}  % mathds macro
\usepackage{authblk}
\usepackage{hyperref}
\usepackage{siunitx}
\usepackage{xcolor}
\usepackage{caption}
\usepackage[utf8]{inputenc}
\usepackage{textcomp}
\usepackage{multirow}

\usepackage{float}
\usepackage{booktabs}
\usepackage{adjustbox}
\usepackage{tikz}
\usetikzlibrary{shapes, arrows, positioning,decorations.markings}
\tikzstyle{vecArrow} = [thick, decoration={markings,mark=at position
	1 with {\arrow[semithick]{open triangle 60}}},
double distance=1.4pt, shorten >= 5.5pt,
preaction = {decorate},
postaction = {draw,line width=1.4pt, white,shorten >= 4.5pt}]
\tikzstyle{innerWhite} = [semithick, white,line width=1.4pt, shorten >= 4.5pt]
\hypersetup{hidelinks}
\usepackage[tmargin=1in,bmargin=1in,lmargin=0.7in,rmargin=0.7in]{geometry}
\newcommand{\pixPos}{\mathcal{X}}
\newcommand{\R}{\mathbb{R}}
\newcommand{\Z}{\mathbb{Z}}
\newcommand{\N}{\mathbb{N}}

\newcommand{\circularity}{\psi}
\newcommand{\axisMajorLength}{v_1}
\newcommand{\radiiRatio}{r}
\newcommand{\minCircumscibedRadius}{r_1}
\newcommand{\perimeter}{\delta}
\newcommand{\feretDiameterMax}{h}
\newcommand{\eccentricity}{e}
\newcommand{\intensityStd}{g_\mathrm{std}} % gray scale
\newcommand{\diameter}{d}
\newcommand{\areaConvex}{a_\mathrm{convex}}
\newcommand{\solidity}{s}
\newcommand{\axisMinorLength}{v_2}
\newcommand{\distTransformStd}{\lambda_\mathrm{std}}
\newcommand{\distTransformMean}{\lambda_\mathrm{mean}}
\newcommand{\maxInscribedRadius}{r_2}
\newcommand{\distTransformMax}{\lambda_\mathrm{max}}
\newcommand{\orientation}{o}
\newcommand{\intensityMean}{g_\mathrm{mean}}
\newcommand{\fractalDimension}{\kappa}
\newcommand{\centroidDeviation}{z}
\newcommand{\intensityMin}{g_\mathrm{min}}
\newcommand{\intensityMax}{g_\mathrm{max}}

\newcommand{\regressionFunction}{\zeta}

\newcommand{\particle}{p_i}
\newcommand{\binaryImage}{\Bar{I}}

\newcommand{\copula}{c}

\newcommand{\nBootstrap}{n_\mathrm{b}}
\newcommand{\apeSolidity}{\mathrm{APE}_s}
\newcommand{\apeSize}{\mathrm{APE}_d}
\newcommand{\density}{f_{\diameter,\solidity}}
\newcommand{\bootstrapDensity}{\widetilde{f}_{\diameter,\solidity}}

\newcommand{\randomForestMapping}{m}
\newcommand{\numberDescriptorsRandomForest}{\nu}
\newcommand{\decisionTree}{T}
\newcommand{\nodeToClassMapping}{w}

\providecommand{\keywords}[1]
{
	{
	\textbf{\textit{Keywords---}} #1
	}
}

\title{Stochastic modeling of particle structures in spray fluidized bed agglomeration using methods from machine learning}
\author{Lukas Fuchs$^{*1}$, Sabrina Weber$^{*1}$, Jialin Men$^2$, Niklas Eiermann$^1$, Orkun Furat$^1$, Andreas Bück$^2$,\\Volker Schmidt$^1$}
\date{}

\begin{document}

\maketitle
\vspace{-3em}
\begin{center}
	\it
$^1$Institute of Stochastics, Ulm University, Ulm, Germany\\
$^2$Institute of Particle Technology, Friedrich-Alexander-Universität Erlangen-Nürnberg, Erlangen, Germany\\
$^{*}$ These authors contributed equally \\
E-mail adresse: lukas.fuchs@uni-ulm.de\\
\end{center}

\begin{abstract}
Agglomeration is an industrially relevant process for the production of bulk materials in which the product properties depend on the morphology of the agglomerates, e.g., on the distribution of size and shape descriptors. Thus, accurate characterization and control of agglomerate
morphologies is essential to ensure high and consistent product quality.
This paper presents a pipeline for image-based inline
agglomerate characterization and prediction of their time-dependent multivariate morphology
distributions within a spray fluidized bed process with transparent glass beads as primary particles. The framework classifies observed objects in image data into three distinct
morphological classes--primary particles, chain-like agglomerates and raspberry-like
agglomerates--using various size and shape descriptors. 
To this end, a fast and robust random forest classifier is trained. Additionally, the fraction of primary particles belonging to each of these classes, either as individual primary particles or as part of a larger structure in the form of chain-like or raspberry-like agglomerates, is described using parametric regression functions. 
Finally, the temporal evolution of bivariate size and shape descriptor distributions of these classes is modeled using low-parametric regression functions and Archimedean copulas.
This approach improves the understanding of agglomerate formation and allows the prediction of process kinetics, facilitating precise control over class fractions and morphology distributions.
\end{abstract}


\section{Introduction}
Agglomeration is a widely used particle formulation process that improves the handling 
of intermediate bulk solids, such as powder, pellets and granules that require further processing before their final application, and generates solid materials with desired end-user product features.
In agglomeration processes, primary particles are combined into larger clusters (agglomerates) by establishing bonds between individual primary particles. These bonds can be established due to interaction forces (e.g., electrostatic or van der Waals forces), by chemical bonds~\cite{D.G.Bika2001,Gluba2001} due to surface reactions of the contacting particles, or by capillary forces due to liquid bridges or by solid bridges. In the latter two examples, a liquid or a solid-containing liquid is required to provide the material that generates the bridges between the primary particles. Agglomeration technologies that utilize these principles are, e.g., pressure agglomeration, binder agglomeration, spray agglomeration and thermal agglomeration~\cite{Wolfgang2002}.
Agglomerates can have superior properties compared to powders of primary particles, e.g., better flowability, higher bulk density and improved mechanical properties, etc., resulting in less dust formation and better strength and durability as well as a high surface-to-volume ratio~\cite{Chris2005,A.K.2018,B.Liu2021}. The 
morphology, i.e., the structure of the agglomerates, 
determines the performance of the bulk material~\cite{Robin2012}.

Spray fluidized bed (SFB) agglomeration is a common agglomeration technique that combines agglomeration and drying in a single vessel by atomizing a binding agent onto a bed of solid particles fluidized by hot gas. SFB agglomeration finds numerous applications in the chemical, food and pharmaceutical industry as it allows mixing, 
structure formation and drying of agglomerates in a single apparatus. 
This enables, e.g., the containment of dust while ensuring cost effectiveness and high efficiency~\cite{ZS.W2019,Fries2014, Stefan2011, Harald1995,Nakarin2008}. These advantages are further enhanced if the spray agglomeration is operated in continuous mode, as this enables uniform operation with constant production rates and uniform agglomerate properties.

In situ process information on the agglomerate structure is required to control the agglomerate formation process with respect to the kinetics and towards autonomous process control. In situ information allows for inverse design of agglomerate structures, i.e., based on product requirements process conditions are selected such that the desired structure is achieved.  Inverse design can be used to implement model-based feedback control that autonomously drives agglomerate formation processes to produce desirable agglomerate structures~\cite{Otto2024,O.A2024,Chen2024, Deng2016, Steven1993}. 

However, agglomeration typically occurs within a timescale of minutes. To enable process control, information on agglomerate formation must be obtained 
on smaller time scales, e.g., within several seconds. An option to fulfill this time constraint are  sequences of in situ high-speed images of individual agglomerates, from which morphological descriptors can be extracted to quantify the current state of the agglomeration process. This is routinely done with respect to agglomerate size and 
some measures of sphericity, more detailed descriptors have not been considered 
yet.
Implementation of the analysis in a recursive fashion, i.e., updating available information by new measurement information, could enhance the real-time capability of structure assessment.

This paper presents a computationally efficient pipeline for image-based particle analysis designed for inline agglomeration characterization for perspective use in autonomous process control, integrating image segmentation, object (particle or agglomerate) classification and parametric modeling of object morphologies. Specifically, the pipeline employs fast, convolution-based denoising~\cite{non_local} combined with Otsu thresholding~\cite{otsu1975threshold} to effectively extract individual objects from image data. A comprehensive set of morphological descriptors is then computed from these segmented objects in order to characterize the agglomeration status and to classify observed objects in image data into three distinct classes: (i) primary particles, which have not yet agglomerated or have broken from agglomerates again; (ii) chain-like agglomerates, consisting of a few primary particles aligned in a nearly linear configuration; and (iii) raspberry-like agglomerates, which denotes large clusters of agglomerated particles with multiple contact points between them.
For fast classification, methods from artificial intelligence~\cite{hastie2005elements} are utilized to achieve high computational efficiency, suitable for inline classification.
The classification enables the subsequent modeling of descriptor distributions for each individual particle/agglomerate classes, using parametric families of 
probability distributions. To gain an even deeper understanding of the state of agglomeration within the process, it is important to consider descriptors that capture both the size and shape of objects within each class. To account for the dependency between these descriptors, we use so-called Archimedean copulas  to determine joint distributions of the considered descriptors ~\cite{nelsen2006introduction, joe2014dependence, furat2019stochastic}. Utilizing regression techniques on the parameters of these copula-based models enables time-dependent modeling and prediction of distributions of size and shape descriptor (vectors), providing insight into structural evolution of agglomerates over time~\cite{weber2021multidimensional}---an essential step towards model-based control, especially in the context of model-predictive control.

This paper is structured as follows. Section~\ref{sec: experimental} describes the experimental study of agglomeration in the SFB using glass beads. Then, the imaging procedure and the subsequent image segmentation is explained in Section~\ref{sec:segmentations}. Section~\ref{sec:descriptors} presents various geometrical descriptors used for classifying particles and agglomerates, with a random forest classifier explained in more detail in Section~\ref{sec: classification}. Moreover, a parametric modeling approach for bivariate distributions of size and shape of particles/agglomerates as well as the temporal evolution of these distributions is explained in Section~\ref{sec: parametric_modeling}. This is followed by a sensitivity analysis that investigates the quantity of model quality for different amounts of data in Section~\ref{sec:sensitivity}. Section~\ref{sec: results} provides the results of classification, time-dependent modeling and the sensitivity analysis. Section~\ref{sec: conclusion} concludes this contribution.

\section{Materials and methods}
\label{sec: materials}
\subsection{Experimental setup}
\label{sec: experimental}
This work is based on the experimental work on SFB agglomeration 
of~\cite{Aisel2024}. We briefly summarize the experimental setup and 
conditions that generate the agglomerates used in this study.

\begin{figure}[h]
     \centering
     \includegraphics[width=.65\linewidth]{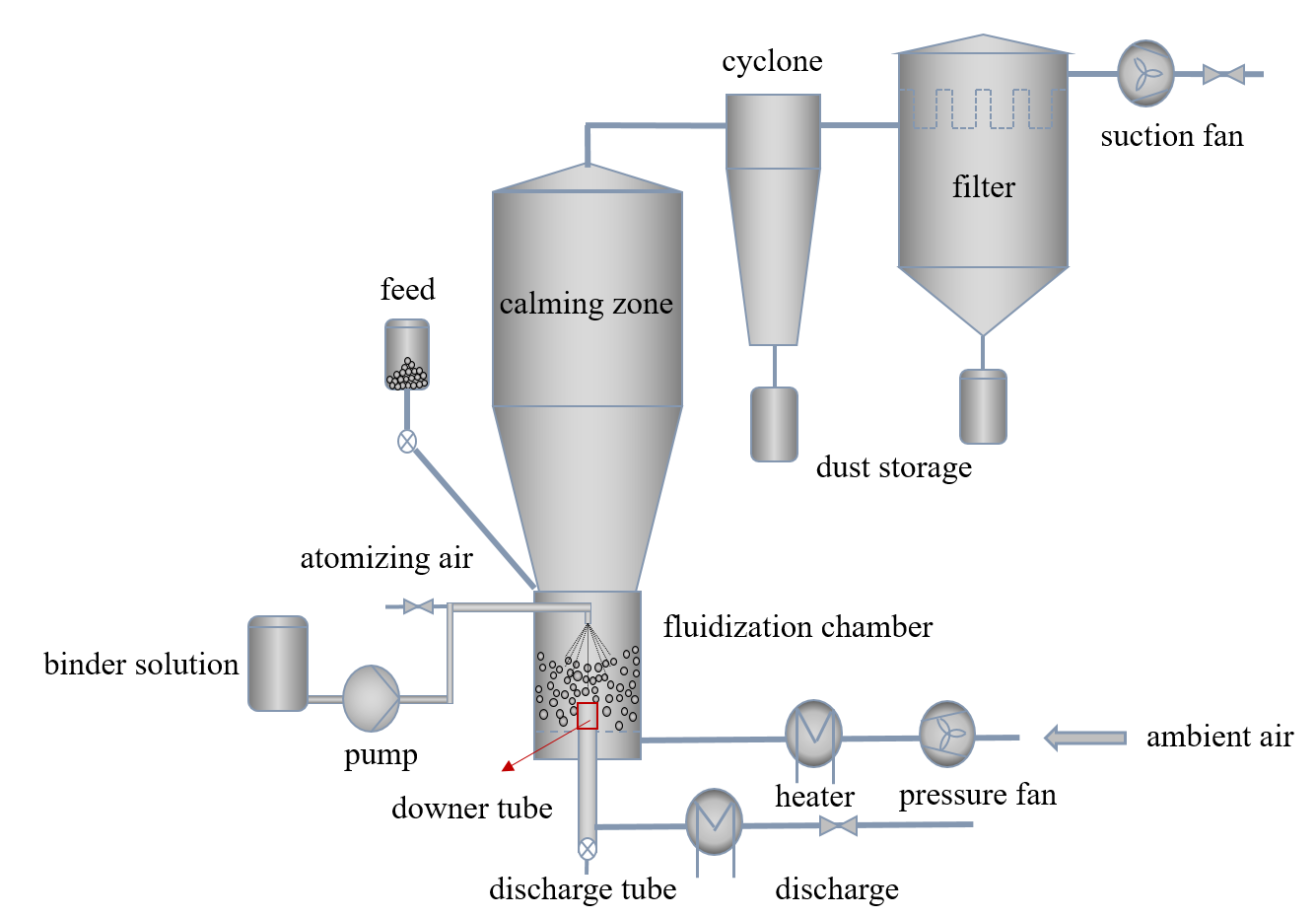}
     \caption{\textbf{Process setup.} Scheme of the pilot-scale spray fluidized bed for continuous spray agglomeration~\cite{Aisel2024}.}
     \label{fig:exp-setup}
 \end{figure}

A pilot scale cylindrical fluidized bed with an inner diameter of \SI{300}{\milli\meter} was 
used (general setup depicted in Figure~\ref{fig:exp-setup}). A two-fluid spray nozzle 
(Düsen-Schlick GmbH, model 940/6 with a hemispheric cap, liquid orifice 
diameter: \SI{0.8}{\milli\meter}) was used, positioned \SI{420}{\milli\meter} above the air distributor plate. The heated
fluidization air enters through the air distributor plate.
Primary particles are suspended by the fluidization gas. An aqueous binder solution is sprayed onto the fluidized particles, so 
that the surface of the particles is wetted and liquid bridges are formed after 
inter-particle collisions. Hot air dries and transforms the liquid bridges into 
solid bridges, thereby forming agglomerates. A filter is used  to 
remove dust from the exhaust gas before it exits the equipment.

In the experiments~\cite{Aisel2024}, transparent glass beads are used as primary particles ($\rho_\mathrm{p}=\SI{2500}{\kilo\gram/\meter^3}$, average volume-weighted diameter $d_{1,3}=\SI{0.24}{\milli\meter}$). Hydroxy-propyl-methyl-cellulose (Pharmacoat 606 from Shin-Etsu, Japan) is taken as binder solution (binder content reported in Table~\ref{tab: experimental}). The experiments were performed in continuous mode with fixed atomization flow rate (\SI{0.08}{\meter^3/\minute}) and average binder spray rate (\SI{32}{\gram/\minute}).  Conditions for all experiments denoted by Experiments~A - E~\cite{Aisel2024} are summarized in Table~\ref{tab: experimental}. Experiments A, B and C were performed with the same binder content (4 wt-$\%$) but different gas inlet temperatures which were set to \SI{80}{\celsius}, \SI{90}{\celsius}, \SI{100}{\celsius}, respectively. Experiments D, B and E were conducted with same gas inlet temperature of \SI{90}{\celsius} and the binder content in the spray was set to 2 wt-\%, 4wt-\% and 6 wt-\%, respectively. All experiments ran for 120 minutes. Samples were taken from the bed and the product outlet at time steps $t\in T=\{10,\ldots,120\}$.
The present work will be based on data collected from their experiments A and E, having the most different experimental conditions (extreme cases).

\begin{table}[]
    \centering
    \caption{\textbf{Process parameters.} Experimental conditions for spray fluidized bed agglomeration}
    \label{tab: experimental}
   \begin{tabular}{|l|c|c|c|c|c|} %'l' aligns the first column to the left; 'c' centers the rest
        \hline
         & A & B & C & D & E \\ \hline
        Inlet gas temperature [℃] & 80 & 90 & 100 & 90 & 90 \\ \hline
        Binder content in wt\% & 4 & 4 & 4 & 2 & 6 \\ \hline
        Fluidization air mass flow rate [kg/hr] & 284  & 286  & 280 & 282 & 280 \\ \hline
        Particle feed rate [g/min] & 145.5 & 158.5 & 181.9 & 165.9 & 154.6 \\ \hline
    \end{tabular}
    
\end{table}

\subsection{Imaging and image processing}\label{sec:segmentations}
In-situ image sequences of particles/agglomerates were acquired using commercial equipment (Camsizer, MicroTracRetsch). There, sample material of the experiment is putted into a dosage hopper. A vibrating chute then guides the particles so that they fall freely in front of an illuminated plane. During this free fall 
high-speed images are taken at 60 fps with an image size of 
1012 pixels by 742 pixels at a  spatial resolution of 
\SI{15}{\micro\meter/pixel}. 
Per experiment and sampling time point, at least 20 images were acquired, with 21.6 objects being observed per image on average .

In order to characterize individual objects within image data and their descriptor distributions, the objects have to be extracted first from the images. 

An image $I\colon \pixPos \to \{0,\ldots,255\}$ is considered as a mapping from the set of all pixel positions $\pixPos =  \{1,\ldots,1012\} \times \{1,\ldots,742\} $  to the set of 8-bit pixel values. To extract individual particles/agglomerates, first each pixel is classified as either background or foreground, i.e., a function $\binaryImage\colon \pixPos \to \{0,1\} $ is computed to classify pixels. More precisely, a pixel $(i,j)\in \pixPos$ with $\binaryImage(i,j)=1$ is considered to be a foreground pixel, whereas  $\binaryImage(i,j)=0$ indicates a background pixel.
The function $\binaryImage$ is computed by utilizing a non-local means filter~\cite{non_local} followed by an Otsu thresholding~\cite{otsu1975threshold}.
The former decreases noise present in an image $I$, whereas the latter is a histogram based heuristic for threshold computation. 
More precisely, by means of non-local denoising a smoothed version $I'\colon \pixPos \to [0,255]$ of $I\colon \pixPos \to \{0,255\}$  is computed which is given by
\begin{align}
I'(x) = \frac{\sum\limits_{z \in Z_x^{21}} w(x, x+z) \cdot I(x+z)}{\sum\limits_{z \in Z_x^{21}} w(x, x+z)},\qquad\text{with}\qquad Z^{21}_x = \{z \in \Z^2 \colon |z|_\infty\leq21, x+z \in \pixPos\},
\end{align}
where $w(x, y)$ is a distance-based weight, given by
\begin{align}
w(x,y) = \exp\!\left(-\frac{1}{100}\sum\limits_{z \in Z^5_{x,y}} \left(I(x+z)-I(y+z)\right)^2\right), \qquad\text{with}\qquad Z^5_{x,y}=\{z \in \Z^2\colon |z|_\infty \leq 5, x+z,y+z \in \pixPos\}.
\end{align}

Here, $|\cdot|_\infty$ denotes the maximum norm and $\Z=\{\ldots,-1,0,1,\ldots\}$ the set of all integers. Intuitively, the weight $w(x, y)$ measures the similarity between the intensity patterns (so-called patches) surrounding the pixels $x$ and $y$. Pixels with more similar local patterns (in terms of the squared sum of their difference) contribute more significantly to the value in the denoised image.

After computing the smoothed image $I'$, Otsu's method is applied to determine a global intensity threshold $\eta\in[0,255]$, see~\cite{otsu1975threshold} for detail. 
By applying the global threshold $\eta$ to $I'$, this yields the phasewise segmentation $\binaryImage\colon \pixPos \to \{0,1\}$, given by
\begin{align}
    \binaryImage(x)=\begin{cases}
        0,& \text{if }I'(x)<\eta,\\
        1, &\text{else}.
    \end{cases}
\end{align}
An object-wise segmentation $p\colon\pixPos \to \N \cup \{0\}$ is subsequently achieved by setting $p(x)=0$ if $\binaryImage(x)=0$
and assigning each connected components of pixel positions $x\in\pixPos$ with $\binaryImage(x)=1$ a unique positive integer. This value is often referred to as the index (or label) of the connected component. Thereby, two pixel positions $x,y\in \pixPos$ are considered to belong to the same connected component if and only if there exist a sequence $z_1,\ldots,z_k \in \{x \in \pixPos \colon \binaryImage(x)=1\}$
such that $x=z_1, z_k=y$ and $|z_i-z_{i+1}|_2=1$ for all $i\in \{1,\ldots,k-1\}$, where $|\cdot|_2$ is the Euclidean norm.  In the following, pixel positions
belonging to the object with index $i$ are denoted by $\particle = \{x \in \pixPos \colon p(x) = i\}$.

 \begin{figure}[H]
     \centering
     \includegraphics[width=0.9\linewidth]{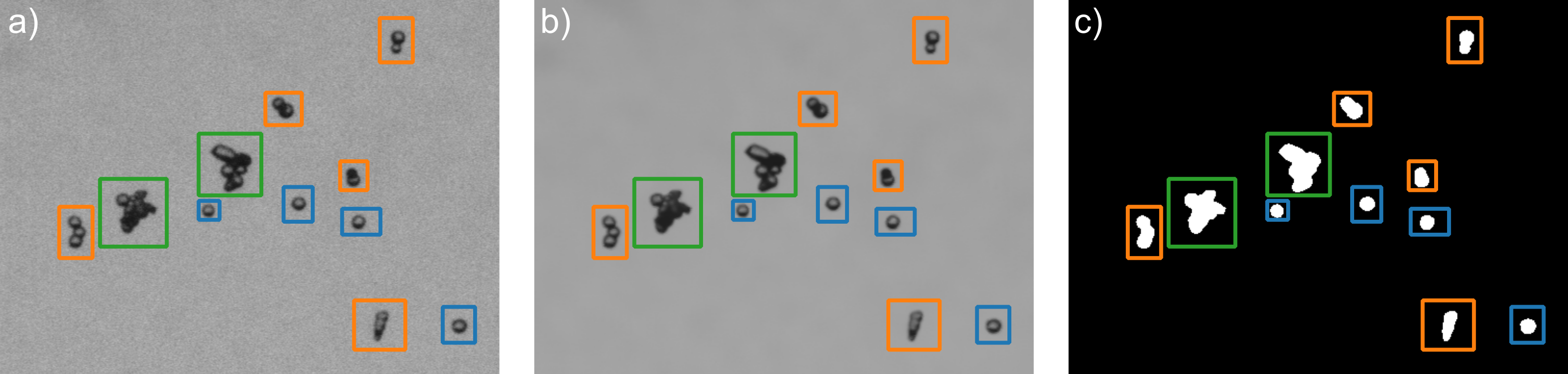}
     \caption{\textbf{Preprocessing of images.} a) Exemplary cutout of an image depicting particles/agglomerates of Experiment A at time step 50 min.
     b) Denoised (non-local means) of a). c) Corresponding phasewise segmentation of b). Exemplary, primary particles are highlighted in blue, chain-like agglomerates in orange and raspberry-like agglomerates in green.}
     \label{fig:segmentation}
 \end{figure}

\subsection{Structure analysis and particle classes}\label{sec:descriptors}

During the SFB agglomeration process, the primary particles agglomerate. Our aim is to characterize the size and shape of both the agglomerates and the primary particles. To quantify the state of agglomeration, we divide the objects observed in image data into three different classes: primary particles, chain-like agglomerates and raspberry-like agglomerates. 

\begin{figure}[H]
    \centering
    \includegraphics[width=0.6\linewidth]{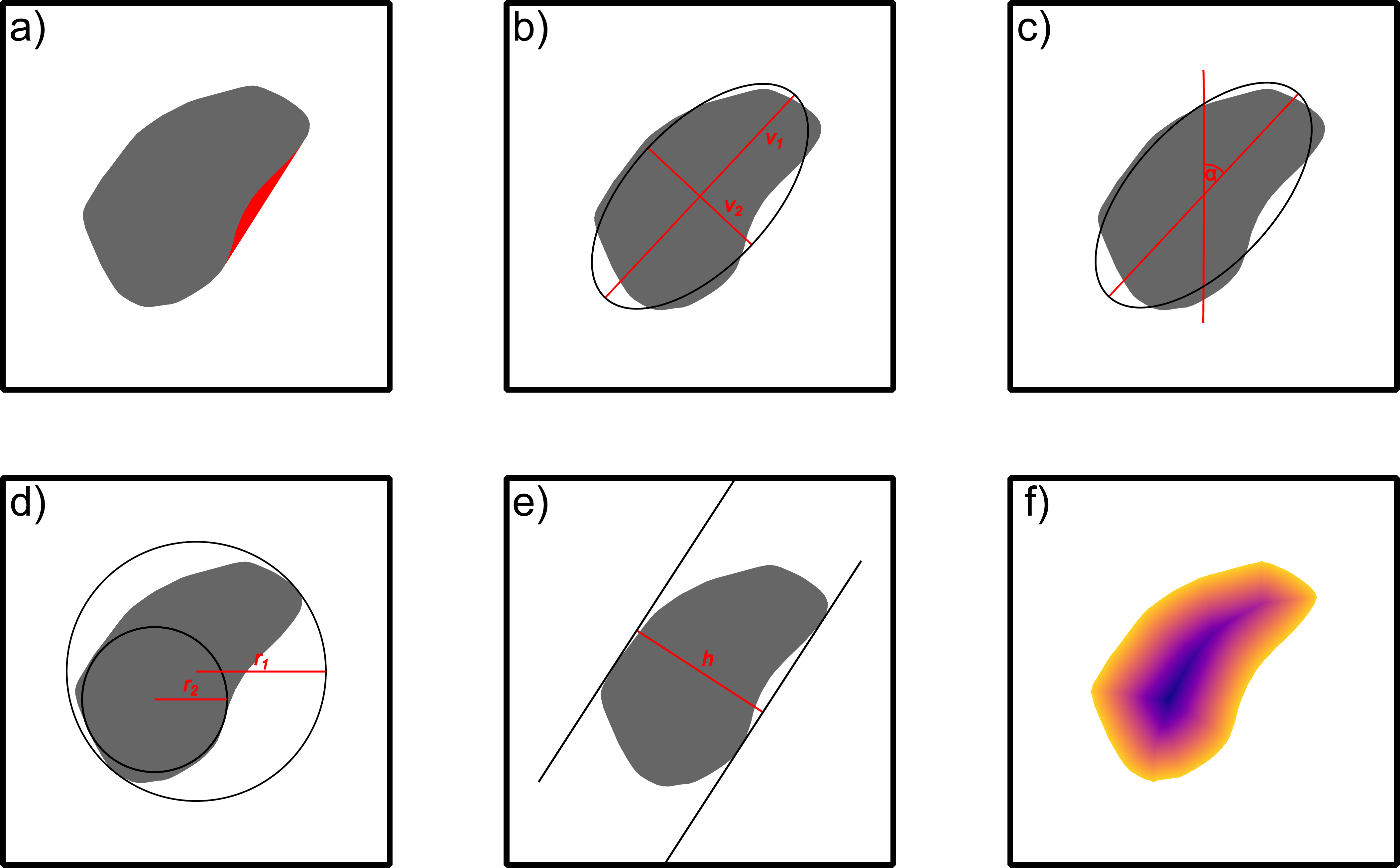}
    \caption{\textbf{Illustration of geometrical descriptors. }(a): The convexity is computed by dividing the area of the object by the area of the convex hull (red). (b): An ellipsoid is fitted to the object and the major and minor axis are determined. (c): The orientation $\alpha$ is given by the angle  between the major axis of a fitted ellipsoid and a vertical line. (d): maximum radius of sphere that inscribe the object and  minimum radius of sphere that enclose the object. (e): distance of each pixel to the object border. (f): One exemplary Feret diameter $h$.}
    \label{fig:descriptors}
\end{figure}

To decide the class membership of objects observed in image data, we use various geometrical descriptors such as the area-equivalent diameter, which is given by $\diameter(\particle) = 2\sqrt {a(p_i)/\pi}$ for particle/agglomerate with index $i$, where $a(p_i)$ is the area of $p_i$. Note that we compute the area of the $i$-th particle/agglomerate $p_i$ by deploying the point-count method ~\cite{chiu2013geometry}. A further geometrical descriptor considered  is the  area $\areaConvex(\particle)$ of the convex hull of $\particle$, where again the point-count method is deployed for the computation of the area. Figure~\ref{fig:descriptors}a) visualizes the difference between the area of $p_i$ and its convex hull in red. 

Then, the solidity of $\particle$ is given by $\solidity(\particle)= a(\particle)/\areaConvex(\particle)$. The solidity is a geometrical descriptor that quantifies how much the shape of $p_i$ deviates from being perfectly convex. A solidity value of $1$ indicates a fully convex object. Additionally, we compute the lengths of the major and minor axes of the ellipse that has the same normalized second central moments as $p_i$, see Figure~\ref{fig:descriptors}b). Further details on the computation of such a ``moment-equivalent'' ellipse can be found in~\cite{pratt2013introduction}. The lengths of the major and minor axes are denoted by $\axisMajorLength(\particle), \axisMinorLength(\particle) \in [0, \infty)$, respectively. Based on these lengths, the eccentricity $\eccentricity(\particle)$ is computed by  
\begin{align}
        \eccentricity(\particle) = \sqrt{1-\begin{pmatrix}\frac{v_2(\particle)}{v_1(\particle)}\end{pmatrix}^2}, 
\end{align}
see~\cite{kenna1959eccentricity}. The eccentricity can be interpreted as a measure that quantifies how much a particle/agglomerate deviates from the circular shape, where a circle has an eccentricity of $0$. The orientation $\orientation(\particle) \in [-\pi/2, \pi/2)$ is computed as the angle between the major axis and the y-axis of the coordinate system, see Figure~\ref{fig:descriptors}c). 
Figure~\ref{fig:descriptors}d), shows the minimal radius $\minCircumscibedRadius(\particle)$ of the sphere that encloses the object  and the maximum radius $\maxInscribedRadius(\particle)$ of the sphere that inscribes the object. The relationship between these two descriptors, i.e., $\radiiRatio(\particle) = \minCircumscibedRadius(\particle)/\maxInscribedRadius(\particle)$, is a further descriptor that is considered for classifying objects.

Additionally, we consider Feret diameters to construct further geometrical descriptors. The Feret diameter into a direction is defined as the distance between two parallel planes that are normal with respect to the chosen direction and enclose the convex hull of an object~\cite{allen2003powder, merkus2009particle}. A visualization of the Feret diameter of an object in an exemplarily chosen direction is given in Figure~\ref{fig:descriptors}e).  For classification purposes, we use the largest Feret diameter of $p_i$ as an additional geometrical descriptor, which we denote by $\feretDiameterMax(\particle)$.

Further geometrical descriptors of $p_i$ are computed from the Euclidean distance transform of $p_i$ which assigns each pixel of $p_i$ with the pixel's Euclidean distance to the boundary of $p_i$, see  Figure~\ref{fig:descriptors}f). Then,  we  compute the mean and standard deviations of the Euclidean distances of all pixels associated with $p_i$, denoted by $\distTransformMean(\particle)$ and $\distTransformStd(\particle)$, respectively. In addition, we compute the maximum Euclidean distance of pixels to the particle border $\distTransformMax(\particle)$ as a geometrical descriptor for classification purposes.
Moreover, for each $p_i$, we compute the perimeter $\perimeter(\particle)$ by identifying boundary pixels using four-neighborhood connectivity. The length of each boundary pixel is then determined based on its local neighborhood configuration: a pixel with two vertically or horizontally adjacent neighbors is assigned a length of $1$, a pixel with two diagonally adjacent neighbors is assigned a length of $\sqrt{2}$, and a pixel with one vertically or horizontally adjacent neighbor and one diagonally adjacent neighbor is assigned a length of $\frac{1+\sqrt{2}}{2}$. The total perimeter $\perimeter(\particle)$ is obtained by summing these lengths and multiplying by the pixel length~\cite{benkrid2000design} is used. With the perimeter and the area of object $i$, we can compute the roundness $\circularity(\particle)$  in the following way, see~\cite{merkus2009particle}:
\begin{align}
    \circularity(\particle) = 4\pi \frac{a(\particle)}{\perimeter(\particle)^2}. 
\end{align}

For classification purposes, we also consider the centroid and the gray value-weighted centroid of $p_i$, denoted as $z_1$ and $z_2$, respectively. The latter is calculated by weighting each pixel's contribution according to its gray value. Then, the Euclidean distance between the centroid and the weighted centroid,  i.e., $\centroidDeviation(\particle) = |z_1(\particle) - z_2(\particle)|$, is considered as additional geometrical descriptor. 

Another descriptor to characterize object $i$ is the fractal dimension $\fractalDimension(\particle)$, which 
is computed using the tiled box counting method, see~\cite{fd1,fd2}. 
Last but not least, the texture of an object is characterized by computing descriptors that quantify the gray values of pixels associated with objects. 

Specifically, we compute the following textural descriptors: the mean gray value $\intensityMean(\particle)$ of pixels associated with $p_i$, the standard deviation $\intensityStd(\particle)$ of gray values of pixels associated with $p_i$, as well as the minimum $\intensityMin(\particle)$ and maximum $\intensityMax(\particle)$ gray value of pixels associated with $p_i$.
In total, this work utilizes $\numberDescriptorsRandomForest=22$ different descriptors for classification purposes. For a better overview, the descriptors are summarized in Table~\ref{tab:geo_desc}.

\begin{table}[H]
    \centering
    \caption{\textbf{Overview of geometrical and textural descriptors for particles.}}
    \label{tab:geo_desc}
    \begin{tabular}{llll}
            \toprule
         Geometrical/textural descriptor & Symbol & Range & Unit  \\
         \midrule
         area-equivalent diameter & $\diameter$ & $(0, \infty)$ & \SI{}{\micro \meter} \\
         area & $a$ & $(0, \infty)$ & $\SI{}{\micro \meter^2}$ \\
         area of convex hull & $\areaConvex$ & $(0, \infty)$ & $\SI{}{\micro \meter ^2}$ \\ 
         solidity & $\solidity$ & $(0,1]$ & - \\
         length of major axis & $\axisMajorLength$ & $(0,\infty)$ & \SI{}{\micro \meter} \\
         length of minor axis & $\axisMinorLength$ & $[0, \infty)$ & \SI{}{\micro \meter}\\
         eccentricity & $\eccentricity$ & $[0,1]$ & - \\
         orientation & $\orientation$  & $[-\pi/2, \pi/2)$ & - \\
         minimal radius of a sphere enclosing the object & $\minCircumscibedRadius$ & $(0, \infty)$ & $\SI{}{\micro \meter}$ \\
         maximum radius of a sphere inscribing the object & $\maxInscribedRadius$ & $(0, \infty)$ & $\SI{}{\micro \meter}$ \\
         ratio of $\minCircumscibedRadius$ and $\maxInscribedRadius$ & $\radiiRatio$ & $(0, \infty)$ & - \\
         largest Feret diameter & $\feretDiameterMax$ & $(0, \infty]$ & $\SI{}{\micro \meter}$ \\
         mean border distance & $\distTransformMean$ & $(0,\infty)$ & \SI{}{\micro \meter} \\
         standard deviation border distance & $\distTransformStd$ & $[0, \infty)$ & \SI{}{\micro \meter} \\
         maximal border distance & $\distTransformMax$ & $(0, \infty)$ & \SI{}{\micro \meter} \\
         perimeter & $\perimeter$ & $[0, \infty)$ & \SI{}{\micro \meter} \\
         roundness & $\circularity$ & $[0, \infty)$ & - \\
         centroid & $z_1$ & $[0,15.3] \times [0, 11.13]$  & mm \\
         gray-value weighted centroid & $z_2$ & $[0,15.3] \times [0,11.13]$  & mm \\
         fractal dimension & $\fractalDimension$ & $[0, 2]$ & - \\
         mean gray value & $\intensityMean$ & $[0, 255]$ & - \\
         standard deviation gray values & $\intensityStd$ & $[0,255]$ & - \\
         minimum gray value & $\intensityMin$ & $[0,255]$ & - \\
         maximum gray value & $\intensityMax$ & $[0,255]$ & - \\
         \bottomrule
    \end{tabular}
    
\end{table}

\subsection{Particle classification}
\label{sec: classification}

In this section we describe the procedure to assign each object in the segmented image data to one of the following classes: primary particles, chain-like agglomerates and raspberry-like
agglomerates. Since hand-labeling is impractical for large datasets, especially in online image processing, and no direct functional relationship is known between geometrical/textural descriptors (as introduced in Section~\ref{sec:descriptors}) and object classes, a classification tool from artificial intelligence is employed. Specifically, a commonly used random forest classification framework~\cite{hastie2005elements} is trained to learn a mapping $\randomForestMapping \colon \R^\numberDescriptorsRandomForest \to \{0,1,2\}$ from the set of interpretable descriptor vectors to a class label, where the label 0 corresponds to primary particles, 1 to chain-like agglomerates, and 2 to raspberry-like agglomerates.

Roughly speaking, a random forest consists of several decision trees that will be used to generate a ``vote'' on the class membership of a particle/agglomerate with descriptor vector $P \in \R^\nu$. The random forest's classification is given by the majority of the votes of the individual trees.

Thus, in order to introduce the notion of a random forest, we first introduce binary decision trees. We will denote a binary decision tree $\decisionTree=(V,E,\mathcal{F},w)$ as a quadruple of a set of vertices $V$, a set of edges $E\subset V\times V$, a set of decision functions $\mathcal{F}$ and a function of $w$ that implicitly assigns a class label to $P$.
The graph  $(V, E)$ forms a perfect binary tree, meaning that each node $v \in V $ has either zero or exactly two children. The set of children of $ v $ is denoted by $ N(v) = \{ v' \in V \mid (v, v') \in E \} $. The tree has a unique root node $ v_\mathrm{r} \in V $, and all leaf nodes $V_\mathrm{l}= \{v \in V \colon N(v) = \emptyset\} $ are at the same depth. The depth of a leaf node $v_\mathrm{l}\in V_\mathrm{l}$ is given by the length $k$ of the shortest sequence $ e_1, \dots, e_k \subset E $ in which consecutive edges share a node, and $v_\mathrm{r},v_\mathrm{l}$  are contained in $e_1,e_k$, respectively.
The set of decision functions $\mathcal{F} = \{f_v\colon \R^\numberDescriptorsRandomForest \to N(v)\colon v\in V, N(v)\neq \emptyset\}$ contains for each non-leaf node a function that maps each descriptor vector $P \in \mathbb{R}^\numberDescriptorsRandomForest$ to a unique child note $u \in N(v)$. 
Furthermore, the function $w\colon V_\mathrm{l} \to \{0,1,2\}$ maps each leaf node to a particle class in $\{0,1,2\}$.
The vote of the decision tree $T$ for a descriptor vector $P$ is determined as follows:
By computing the unique path $(v_\mathrm{r}, f_v(P), f_{f_v(P)}(P), \ldots, v_\mathrm{l}) \subset V$ from the root node $v_\mathrm{r}$ of the tree $T$ to a leaf node $v_\mathrm{l}\in V_\mathrm{l}$ is identified. Then, the tree $T$ assigns the object with the descriptor vector $P$ to the class with label $\nodeToClassMapping(v_\mathrm{l})$.

In this work, the decision functions $f_v\in \mathcal{F}$ are threshold-based decisions applied to individual descriptors. Specifically, the decision functions take the form:
\begin{align}
    f_v(P)=\begin{cases}
        v_1,& \text{if } P_j >P^*,\\
        v_2,& \text{else},
    \end{cases}
\end{align}
where $\{v_1, v_2\}=N(v)$ are the child nodes of $v$, $P^* \in \mathbb{R}$ is a threshold and $P_j$ is the $j$-th descriptor within the descriptor vector $P=(P_1,\ldots,P_\nu)$. Thus, the decision function $f_v$ is uniquely determined by the tuple $(j,P^*)$ of the considered descriptor given by $j\in\{1,\ldots,\numberDescriptorsRandomForest\}$ of the descriptor vector $P\in \R^\numberDescriptorsRandomForest$ and the corresponding threshold $P^*$.

For training data consisting of a sequence $\mathcal{P}\subset \R^\numberDescriptorsRandomForest$ of  descriptor vectors and corresponding class labels in $\{0,1,2\}$ for each descriptor vector, a decision tree $T$ can be efficiently computed by the so-called CART algorithm~\cite{hastie2005elements}. Specifically, for a perfect binary tree $(V,E)$ this algorithm, iteratively, determines  the functions $f_v \colon \mathbb{R}^\numberDescriptorsRandomForest \to N(v)$ for all $v\in V$ with $N(v)\neq \emptyset$  by greedily maximizing some quality measure $Q(f_v,\mathcal{P})$ for some set of descriptor vectors $\mathcal{P}$. 

We consider a quality measure that is based on the so-called Gini coefficient $G$. The Gini coefficient gives a measure for the impurity of classes of some sequence $\mathcal{P}$ of descriptor vectors. More precisely, this coefficient is given by
\begin{align}
    G(\mathcal{P})=\sum_{k=0}^2 (1-q_k)q_k, 
\end{align}
where $q_k\in [0,1]$ is the fraction of descriptor vectors of  $\mathcal{P}$ that were assigned to the class $k$.
The quality $Q(f_v,\mathcal{P})$ of a function $f_v$ for a sequence $\mathcal{P}\subset \R^\numberDescriptorsRandomForest$ of descriptor vectors is then given by
\begin{align}
    Q(f_v,\mathcal{P})= \sum_{u \in N(v)} G(\mathcal{P}_u)  |\mathcal{P}_u|,
\end{align}
where $\mathcal{P}_u$ is the maximum subsequence of  $P$ for which it holds that $f_v(P)=u$ for all $v\in \mathcal{P}_u$, and $|\cdot|$ denotes the length of the sequence under consideration.

The CART-algorithm starts with the root $v_\mathrm{r}\in V$ of a tree $(V,E)$ and the sequence $\mathcal{P}$ of all measured particle descriptors, and computes a function $f_{v_\mathrm{r}}$ that minimizes $Q(f_{v_\mathrm{r}},\mathcal{P})$. Afterward, the same procedures are repeated for child nodes $u\in N(v_\mathrm{r})$ with the corresponding sequences $\mathcal{P}_u$ of descriptor vectors until, for all non-leaf nodes $v\in V$, the function $f_v$ is determined.
At a last step, the value of the function $\nodeToClassMapping\colon \{v\in V\colon N(v)=\emptyset\} \to \{0,1,2\}$ is set to the class in $\{0,1,2\}$ that is most frequently among the classes of the particle descriptor vectors  of $\mathcal{P}_v$ in the training data. See~\cite{hastie2005elements} for more details.

This procedure results in an optimal classification of the training data~\cite{hastie2005elements}; however, considering all components $P_j,~j\in \{1,\ldots,\numberDescriptorsRandomForest\}$  of the descriptor vectors $P\in\mathcal{P}$ in the computation of $f_v,~v\in V,~N(v)=\emptyset$ often leads to overfitting, and thus poor performance on unseen data.
To mitigate this, a subset $J\subset \{1,\ldots,\numberDescriptorsRandomForest\}$ of $|J|<\numberDescriptorsRandomForest$ descriptor component indices is chosen before training a decision tree $T$. Then, when computing, the decision functions $\mathcal{F}$ are restricted to descriptors with an index in $J$. 

A random forest is a collection of $B\in \N$ binary decision trees $\decisionTree_1, \ldots, \decisionTree_B$. 
These are constructed by independently calibrating $B>1$ binary decision trees $T_1,\ldots,T_B$, each of which  with an independently chosen random subset  $J_1,\ldots,J_B\subset \{1,\ldots,\numberDescriptorsRandomForest\}$ of descriptor indices. Then, the random forest can be deployed for classifying an object with descriptor vector $P\in\R^\nu$, by (i) determining the $B$ class assignments of $P$ according to decision trees $T_1,\ldots,T_B$ followed by (ii)  majority voting along these class assignments. This approach effectively addresses the overfitting problem. 
The choice of a random tree's hyperparameters---the  number $B$ of considered decision trees as well as the number $|J|$ of considered particle descriptors, and the depth of the binary trees $(V_1,E_1),\ldots,(V_B,E_B)$---is subject to the training procedure. For more details and heuristics for choosing these hyperparameters, see~\cite{zhang2012ensemble}.
We optimize these hyperparameters, by means of a grid-search. The grid-search is performed by training and evaluating random forests with 50, 100 and 200 decision trees, considering maximum tree depths of 3, 5, 7 and $\numberDescriptorsRandomForest,$ and sizes of $J$ equal to $5 \approx \sqrt{\numberDescriptorsRandomForest}$ and $\numberDescriptorsRandomForest$.
During the gird-search, all possible combinations of hyperparameters are used for training. The best hyperparameters are those that lead to the random forest with the highest percent of correctly classified particles/agglomerates on the training data.
The prediction of a class label using a random forest, is very fast, thus combined with the fast computation of the image segmentation and particle descriptors, an online classification of imaged particles/agglomerates is enabled. Furthermore, the decisions $f_v$ of the individual trees can be interpreted, giving insight into, the influence of interpretable particle descriptors on the classification, see Section~\ref{sec:results_classification}.

\subsection{Particle class and descriptor modeling}\label{sec:copula}

The previously introduced automatic classification method enables the efficient classification of a large particle/agglomerate database. In this section, the focus is on modeling the evolution of primary particles, chain-like and raspberry-like agglomerates over time. This is achieved by analyzing two aspects: first, the temporal evolution of the class size fractions, and second, the evolution of bivariate distributions of descriptors within individual classes over time.
In the following, methods for modeling these evolutions are presented generally. Later, these methods will be deployed to data sets acquired from the Experiments A and E, see Section~\ref{sec: results} for further details.

\subsubsection{Temporal evolution of class sizes}
\label{sec: class_sizes}
First, we analyze the size of an object class over time by means of the area-weighted fraction of objects belonging to this class compared to all observations. For this, let $M$ be a set of pairs $(t,y)$ of time steps $t\in T$ and corresponding size fractions $y\in[0,1]$ of the considered class. We will model the value of $y$ with respect to $t$ by means of a parametric regression function $\regressionFunction \colon [0, \infty) \to \R$.
The specific form of the regression functions is given by
\begin{align}
    \regressionFunction(t) = c_1-c_2\, \exp\!{(-{c_3}\, t)}
    \label{eq:regressionfunction}
\end{align}
for any $t \in [0,120]$, where $c_1,c_2,c_3\in \R$ are the parameters of the regression function and $\exp\colon \R \to [0, \infty)$ is the Euler's function. Thereby, the parameter $c_1$ describes the asymptotic value of $g$ for $t\to \infty$; $c_2$ determines the value of $g$ for small values of $t$, and $c_3$ determines how fast $g$ reaches its asymptotic value $c_1$. 

For a set $M$ of pairs $(t,y)$ the parameters of the regression function $\regressionFunction$ can be calibrated by minimizing the mean squared error (MSE),  i.e., the values of $(c_1,c_2,c_3)\in \R^3$ are given by
\begin{align}
    (c_1,c_2,c_3) = \underset{(c_1,c_2,c_3)\in \R^3}{\mathrm{argmin}} \frac{1}{|M|} \sum\limits_{(t,y)\in M} (\regressionFunction(t)-y)^2.\label{eq: argmin_regression}
\end{align}

For both experiments, Experiment A and Experiment E, we use the regression function, given by Equation~\eqref{eq:regressionfunction}, to model the size fractions of the primary particles, and to model the size fractions of the raspberry-like agglomerates over time. Thus in these four cases, the set $M$ consist of pairs $(t,y)$ of a time steps $t\in T$ and the respective area-weighted size fractions of primary particles (or raspberry-like agglomerates) at time $t$ at all respective experiments. 
In order to ensure that the size fractions, modeled by means of the resulting regression functions, are fractions, the evolution of the size fraction of the chain-like agglomerate class is modeled by $1-g_1-g_2$, where $g_1,g_2$ are the respective regression functions of the primary particles and raspberry-like agglomerates.
Although there exist values of $(c_1,c_2,c_3)\in \R^3$ for which $g_1(t),g_2(t)\notin [0,1]$ for $t\in [10,120]$, this does not occur in the application, see Section~\ref{sec:results_regression_size}.  The choice of modeling the size fraction of the chain-like agglomerates as the complement of the other two classes is intuitive, since chain-like agglomerates appear as an intermediate class between non-agglomerated primary particles and the desired final product, the raspberry-like agglomerates.
In Section~\ref{sec: results} the results of the regressions for both considered experiments and all classes of objects are shown.

\subsubsection{Parametric copula-based modeling}
\label{sec: parametric_modeling}
To gain a more comprehensive understanding of the agglomeration process, the bivariate probability distributions of area-equivalent diameter $\diameter$ and solidity $\solidity$ within the three classes are modeled parametrically for each time step.  This parametric modeling facilitates the time-dependent regression of probability densities by performing the regression in a lower dimensional space instead, namely, on the set of model parameters. Consequently, it enables the regression of particle descriptor distributions and thus allows for predicting these distributions for time steps which were not directly measured. In this manner a temporal model for the bivariate probability distribution of descriptor vectors can be obtained. This modeling approach is deployed for each class and observed in  Experiments A and E individually, see  Section~\ref{sec: results}. However, to facilitate the notation the methodology is introduced in general for one and two-dimensional vector data in this section. Therefore, unless stated otherwise, we consider a single class of objects observed in a single experiment in the remainder of Section~\ref{sec: parametric_modeling}.

\paragraph{Univariate densities.} 
First, we want to determine the univariate densities $f_\diameter, f_\solidity: \R \to [0, \infty)$ and the corresponding cumulative distribution functions $F_\diameter, F_\solidity: \R \to [0,1]$ of 
 $\diameter$ and  $\solidity$ for an individual time step.
Therefore, we use the parametric families 
$\mathcal{G} = \{ \text{normal}, \text{log-normal}, \text{gamma}\}$, which are further specified in Table~\ref{tab: univariate_dist} and in~\cite{johnson1994continuous}. However, note that the solidity takes values in $[0,1]$ and the area-equivalent diameter takes values in $(0, \infty)$. Thus, for parametric families with different support we consider truncated versions to ensure a correct support. To determine suitable parameter values and the best family, maximum likelihood estimation is used~\cite{aitkin2010statistical}. In this manner, we obtain for each family $G_\diameter, G_\solidity \in \mathcal{G}$
parametric probability densities $f_\diameter^{G_\diameter, \omega_{t,d}}, f_\solidity^{G_\solidity, \omega_{t,s}}: \R \to [0, \infty)$ of $\diameter$ and  $\solidity$ at the considered time step $t$ as well as their corresponding distribution functions $F_\diameter^{G_\diameter, \omega_{t,d}}, F_\solidity^{G_\solidity, \omega_{t,s}} \colon \R \to [0,1]$. Note that $\Omega_{G_\diameter}, \Omega_{G_\solidity}$ denotes the parameter space of family $G_\diameter$ and $G_\solidity$. The used parameter vector for $\diameter$ and $\solidity$ are denoted by $\omega_{t,\diameter} \in \Omega_{G_\diameter}$ and $\omega_{t,\solidity} \in \Omega_{G_\solidity}$ denotes. 

More details on fitting these parameters vector and determining the overall best family for all time steps is given in subsequent paragraphs.

\begin{table}[H]
    \centering
    \caption{\textbf{Parametric univariate distributions.} Parametric families of univariate distributions with corresponding density, support and parameter space $\Omega$,
    where  $\Gamma$ denotes the gamma function~\cite{johnson1994continuous}.}
    \label{tab: univariate_dist}
    \begin{tabular}{c|c|c|c|c}
        parametric family & probability density & support &  $\omega \in \Omega$  \\ \toprule
        normal &  $\frac{1}{\sqrt{2\pi\sigma^2}}e^{-(x-\mu)^2/(2\sigma^2)}$ & $(-\infty, \infty)$ & $(\mu , \sigma) \in \R \times (0, \infty)$ \\
        log-normal  & $\frac{1}{x\sqrt{2\pi\sigma^2}}e^{-(\log^2(x)-\log^2(\mu))/(2\sigma^2)}$ & $(0, \infty)$ & $(\mu, \sigma) \in (0, \infty) \times (0, \infty)$ \\
        gamma  & $\frac{1}{\beta^\alpha} x^{\alpha-1}e^{\frac{1}{\beta}x}\frac{1}{\Gamma(\alpha)}$ & $[0, \infty)$ & $(\alpha, \beta) \in (0, \infty) \times (0, \infty)$
    \end{tabular}  
\end{table}

\paragraph{Bivariate copula-based densities.} To describe the structure of objects at an individual time step, the interdependence between the size and shape descriptors can be modeled. To do so, the bivariate distribution of solidity and area-equivalent diameter is modeled using copulas, which
provide a flexible approach by decoupling the marginal distributions from their dependence structure. More precisely, a function $C: [0,1]^2 \to [0,1]$ is called a bivariate copula if $C$ is the cumulative distribution function of a two-dimensional random vector with standard uniformly distributed marginals. Then, according to Sklar's representation formula (see~\cite{nelsen2006introduction}), for the joint cumulative distribution function $F_{\diameter,\solidity}: \R^2 \to [0,1]$ of $\diameter$ and $\solidity$, there exists a copula $C$ such that
\begin{align}
\label{skl.rep.for}    F_{\diameter,\solidity}(x_\diameter,x_\solidity) = C(F_\diameter(x_\diameter),F_\solidity(x_\solidity)),  
\end{align}
for $x_\diameter \in [0, \infty)$ and $x_\solidity \in [0,1]$ denoting the particle descriptor values. 

Note that, assuming that $F_\mathrm{\solidity,\diameter}$ and $C$ are differentiable, it follows from Equation~\eqref {skl.rep.for}
that 
the joint probability density $f_{\diameter,\solidity}: \R^2 \to [0, \infty)$ of $\diameter$ and $\solidity$ is given by
\begin{align}
f_{\diameter,\solidity}(x_\diameter,x_\solidity) = c(F_\diameter(x_\diameter),F_\solidity(x_\solidity))f_\diameter(x_\diameter)\,f_\solidity(x_\solidity),
    \label{eq: bivariate_desnity}
\end{align}
where $c: [0,1]^2 \to [0, \infty)$ is the probability density  of $C$. Thus, in this case, the differential version of Sklar's representation formula given in Equation~\eqref{eq: bivariate_desnity}
can be used to construct bivariate densities by fitting the univariate margins (see the preceding paragraph) and then a bivariate copula.

For modeling bivariate distributions, so-called Archimedean copulas are used in the present paper. The definition of these copulas is based on Archimedean generators $\varphi:[0,1] \to [0, \infty)$, which are continuous, strictly decreasing functions such that $\varphi(1) = 0$. Moreover, let $\varphi^{[-1]}:[0, \infty) \to [0,1]$ be the pseudo inverse of $\varphi$, i.e., $\varphi^{[-1]}(x) = \varphi^{-1}(x)$ if $0 \leq x \leq \varphi(0)$ and $\varphi(x) = 0$ if $x \geq \varphi(0)$, where $\varphi^{-1}$ denotes the inverse of $\varphi$. An Archimedean copula is then given by 
\begin{align}
    C(u_1, u_2) = \varphi^{[-1]}(\varphi(u_1)+\varphi(u_2)),
\end{align}
for any $u_1, u_2 \in [0,1]$, see e.g.~\cite{nelsen2006introduction}. To model bivariate densities, various parametric families $\{\varphi_\theta: \theta \in \Theta\}$ of Archimedean generators are considered, see Table~\ref{tab:archimedean}. The space of admissible parameters is denoted by $\Theta \subset \R$. Each family of Archimedean generators leads to a parametric family of copula densities $\{c_\theta : \theta \in \Theta\}$ given by 

\begin{align}
    c_\theta(u_1,u_2) = \frac{\partial^2}{\partial u_1 \partial u_2} \varphi_\theta^{[-1]}(\varphi(u_1)+\varphi(u_2))
\end{align}
for any $u_1,u_2 \in [0,1]$ and $\theta \in \Theta$. 

Even further families of copula densities can be constructed, by considering rotating copula densities within a given family by multiples of $90^\circ$. More precisely, $c_\theta$ can be rotated around the midpoint $(0.5,0.5)$ by $90^\circ, 180^\circ$ or $270^\circ$ to obtain copula families. To determine the optimal copula family and density parameter, maximum likelihood estimation is used as in the case of fitting the univariate distributions.

For the parametric case with Archimedean copulas we adapt Equation~\eqref{eq: bivariate_desnity} in the following. Therefore, we consider a set $\mathcal{Z} = \{\text{Frank, Joe, Clayton, Gumbel, Ali-Mikhail-Haq\}}$ of copula types, each of which induces a parametric family of copula densities. The bivariate density with previously fitted marginal distributions and the copula density $c^{Z,\theta_t}_\mathrm{\diameter, \solidity}:[0,1]^2 \to \R$ of family $Z$ with parameter $\theta_t \in \Theta_Z \subset \R $, where $\Theta_Z$ is the parameter space of the copula family $Z$ defined in Table~\ref{tab:archimedean}, is then given by 
\begin{align}
    f_{{\diameter,\solidity}}^{Z, \theta_t}(x_\diameter, x_\solidity) = c^{Z, \theta_t}_\mathrm{\diameter, \solidity}(F^{G_\diameter, \omega_{t,\diameter}}_{\diameter}(x_\diameter), F^{G_\solidity, \omega_{t,\solidity}}_{\solidity}(x_\solidity))f^{G_\diameter, \omega_{t,\diameter}}_{\diameter}(x_\diameter) f^{G_\solidity, \omega_{t,\solidity}}_{\solidity}(x_\solidity),
    \label{eq: bivariate_density_parametric}
\end{align}
for any $x_\diameter \in [0,\infty)$ and $x_\solidity \in [0,1]$.

\begin{table}[H]
\centering
\caption{\textbf{Archimedean generators. }Parametric families $\{\phi_\theta: \theta \in \Theta\}$ of Archimedean generators, together with their set of parameters $\Theta \subset \R$.}    \label{tab:archimedean}
    \begin{tabular}{ >\centering m{1.5cm} | >\centering m{2.5cm} | >\centering m{2.5cm}| >\centering m{2.5cm} | >\centering m{2.5cm} | >\centering m{2.5cm}}
         copula & Frank & Joe & Clayton & Gumbel & Ali-Mikhail-Haq \cr \toprule
         $\varphi_\theta (u)$ & $-\ln \frac{\exp(-\theta u)-1}{\exp(-\theta)-1}$ & $-\ln(1-(1-u)^\theta)$ & $\frac{1}{\theta}(u ^{-\theta} -1)$ & $(- \ln u)^\theta$  &  $\ln \frac{1-\theta(1-u)}{u}$ \cr
          $\Theta$ & $\R \setminus \{0\}$ & $[1, \infty)$ & $(0, \infty)$ & $[1, \infty)$ & $[-1,1)$
    \end{tabular}
    
\end{table}

\paragraph{Time-dependent regression of distribution parameters.} For our purpose of a time-dependent regression, it has to be ensured that the same parametric family of probability densities is deployed to model the distribution of a descriptor (vector) within a class for all time steps  $T = \{10,20,30,40,50,60,70,80,90,100,110,120\}$ minutes of an experiment. Recall that, we explain the procedure for a single class of objects observed in one single experiment to simplify the notation. For example, we can consider raspberry-like agglomerates in Experiment A. In all other cases the procedure is analog. We consider first the set $\mathcal{G} = \{\text{norm, lognorm, gamma}\}$ of distribution types, i.e., each type in $\mathcal{G}$ has an associated parametric family of univariate probability densities. In order to identify a parametric family of probability densities   
 that can adequately model the the probability density of $\diameter$ for all time steps, we identify the optimal distribution type $\widehat{{G}}_\diameter \in \mathcal{G}$  that maximizes the overall likelihood, i.e., the distribution type is given by
 \begin{align}
         \widehat{G}_\diameter = \underset{G \in \mathcal{G}}{\arg \max} \sum_{t \in T} \left( \underset{\omega_{t,d} \in \Omega_G}{\max} \sum_{x \in D_{t, \diameter}} \log (f^{G,\omega_{t,d}}_{d}(x)), \right),\label{eq:likelyhood}
 \end{align}
where $D_{t, \diameter}$ denotes the set of descriptors $\diameter$ observed at time step $t$. The parametric family of univariate probability densities of the chosen distribution type $\widehat{G}_\diameter$ is then fitted to the data $D_{t, \diameter}$ for each time step by means of maximum likelihood estimation. Thus, we obtain a sequence of fitted parameters $\widehat{\omega}_{t,d} \in \Omega_{\widehat{G}_\diameter}$ for each time step $t\in T$. In other words, for each $t\in T$ we obtain a sequence of univariate probability densities $f^{\widehat{G}_\diameter,\widehat{\omega}_{t,d}}_{d}$ of $\diameter$ that all stem from the same parametric family. 
Analogously, univariate probability densities $f^{\widehat{G}_s,\widehat{\omega}_{t,s}}_{s}$ can be determined for modeling the distribution of the solidity for each time step $t\in T$ with the same parametric family.

After fitting the univariate distributions of both descriptors, we  determine the best parametric family of copula densities by using Equation~\eqref{eq: bivariate_density_parametric}. More precisley, to chose the best-fitting copula family, the maximized likelihood values for the copula with previously fitted marginal distributions are summed up as above to identify the best-fitting family. More precisely, the best copula is provided by
\begin{align}
    \widehat{Z} = \underset{Z \in \mathcal{Z}}{\text{arg max}} \sum_{t \in T} \left(\max_{\theta_t\in \Theta_Z} \sum_{(x_\diameter,x_\solidity) \in D_{t,(\diameter,\solidity)}} \log(f_{\diameter,s}^{Z,\theta_t}(x_\diameter,x_\solidity)) \right),
    \label{eq: likelyhood_biv}
\end{align}
where  $D_{t,(\text{\diameter, \solidity})}$ is the set of two-dimensional descriptor vectors (pairs of area-equivalent diameter and the solidity) measured at time $t\in T$.  Moreover, we obtain a sequence of fitted parameters $\widehat{\theta}_t \in \Theta_{\widehat{Z}}$ for each time step $t \in T$, i.e., we obtain a sequence of copulas $c_\mathrm{\diameter, \solidity}^{\widehat{Z}, \widehat{\theta}_t}$, which together with the univariate distributions define the bivariate density $f_\mathrm{\diameter, \solidity}^{\widehat{Z},\widehat{\theta}_t}$, see Equation~\eqref{eq: bivariate_density_parametric}.

After this procedure we have identified a suitable low-parametric model for describing the temporally resolved bivariate distribution of area-equivalent diameters and solidity and the parameters, i.e., the vector $\tau_t = (\widehat{\omega}_{t,\text{\diameter}}, \widehat{\omega}_{t,\text{\solidity}},\widehat{\theta_t}) \in \Omega_{\widehat{G}_\text{\diameter}} \times \Omega_{\widehat{G}_\text{\solidity}} \times \Theta_{\widehat{Z}}$, which is fitted to the set of two-dimensional descriptor vectors for each particle class, experiment and time step $t$.

At this point we are able to describe the bivariate distribution of area-equivalent diameter and the solidity for all particle classes and experiments at $t\in T=\{10,20,\ldots,120\}$. However, it is of interest to predict these distributions for all $t\in [10,120]$. For this purpose, a regression of the parameters of the copula model is utilized. More precisely, the regression function of Equation~\eqref{eq:regressionfunction} is utilized to predict the vector $\tau_t$ for all $t\in [10,120]$.
Note that, for each time step a different number of observed agglomerates is available. In order to always weight each available data point equally, the regression curve is fitted by minimizing the weighted MSE. Thus, Equation~\eqref{eq: argmin_regression} is adjusted as follows
\begin{align}
    (c_1,c_2,c_3) = \underset{(c_1,c_2,c_3) \in \R^3}{\text{argmin}} \sum_{t \in T} ((\tau_t^i - \ \regressionFunction(t)) |D_t|)^2,
    \label{eq: loss_paramter_fitting}
\end{align}
where $\tau_t^i$ is the $i$-th entry of $\tau_t$ and $|D_t|$ is the amount of measured data at time step $t$ . By utilizing a regression of the five parameter values over time by the described procedure, statements can be made about the distributions of unobserved points in time. Moreover, fitting functions to the parameter course over time allow to make predictive statements on the parametric distributions. Note that the described method is applied for Experiments A and E as well as for each particle class, i.e., primary particle, chain-like agglomerates and raspberry-like agglomerates. The results are presented in Section~\ref{sec: result_descripotr} below.

\subsection{Sensitivity analysis of fitting procedure of bivariate distributions}\label{sec:sensitivity}
The present study is based on a large data set of descriptors computed from experimentally measured image data. However, the question arises, how sensitive the presented procedure is to the amount of available data, and consequently, how many measurements are necessary in order to achieve a reasonable quality of fit. To answer these questions, a bootstrap sampling-based sensitivity analysis of the presented modeling approach is deployed~\cite{good2006resampling}.
This involves the quantitative analysis of model fits that are achieved on a data set containing only a fraction of the measured data, allowing for the analysis of the added value of an increasing amount of available data. 

To analyze the sensitivity of the fit of the probability density $f_{\diameter,\solidity}$ that is based on some data $\mathcal{Y}$ (e.g., $\mathcal{Y}=D_{t,(\text{\diameter, \solidity})}$ for some $t \in T$), first, a bootstrap sample $\widetilde{\mathcal{Y}}$
of size $\nBootstrap\in\N$ is constructed by drawing $\nBootstrap>0$ data points uniformly at random from the set $\mathcal{Y}$. Then, a second probability density $\widetilde{f}_{\diameter,\solidity}$ is fitted with the data in $\widetilde{\mathcal{Y}}$. In this manner, we can investigate the discrepancy (see below for further details) of $f_{\diameter,\solidity}$ to a fit $\widetilde{f}_{\diameter,\solidity}$ that has been achieved with fewer data.

In the present paper, the discrepancy between three probability densities $\bootstrapDensity$ and $\density$ is quantified in two ways. First, the absolute percentage errors $\apeSize$ and $\apeSolidity$ of the  expected values of the marginal distributions are considered this is given by
\begin{align}
    \apeSize(\density,\bootstrapDensity)= \frac{|\int_0^\infty x (\widetilde{f}_d(x)-f_d(x))\, dx|}{|\int_0^\infty x f_d(x)\, dx|},\label{eq:apeds}\\
    \apeSolidity(\density,\bootstrapDensity)= \frac{|\int_0^\infty x (\widetilde{f}_s(x)-f_s(x))\, dx|}{|\int_0^\infty x f_s(x)\, dx|},
    \label{eq:apes}
\end{align}
where $f_d,f_s\colon \R \to [0,\infty)$ and $\widetilde{f}_d,\widetilde{f}_s\colon \R \to [0, \infty)$ are the marginal probability densities of $\density$ and $\bootstrapDensity$, respectively. 
In order to quantify not only the discrepancy of the marginal distributions, but also the discrepancy of the dependency structures of the marginal distributions, a second measure $L(\density,\bootstrapDensity)\in [0,2]$, is utilized. This measure compares the copula densities $\copula, \widetilde{\copula}\colon[0,1]^2\to [0,\infty)$, of $\density,\bootstrapDensity$ by means of the  $L_1$-norm and is given by
\begin{align}
L(\density,\bootstrapDensity) &= \int\limits_{0}^1\int\limits_{0}^1 \left| \copula(x, y) - \widetilde{\copula}(x, y) \right| \, dx \, dy.\label{eq:copula_loss}
\end{align}
A value of $L(\density,\bootstrapDensity)$ close to zero, corresponds to a high similarity, whereas a value close to two, corresponds to extreme dissimilarity.

By means of the outlined bootstrapping approach, we can investigate the goodness of fit in dependence of the number $\nBootstrap$ of sampled data points. In other words, this approach enables us to assess the number of objects and, consequently, the number of measurements necessary to achieve the desired precision in our model fits; see Section~\ref{sec: results_Sensitivity}.

\section{Results and discussion}
\label{sec: results}
\subsection{Segmentation}
\begin{figure}[H]
    \centering
    \includegraphics[width=0.35\linewidth]{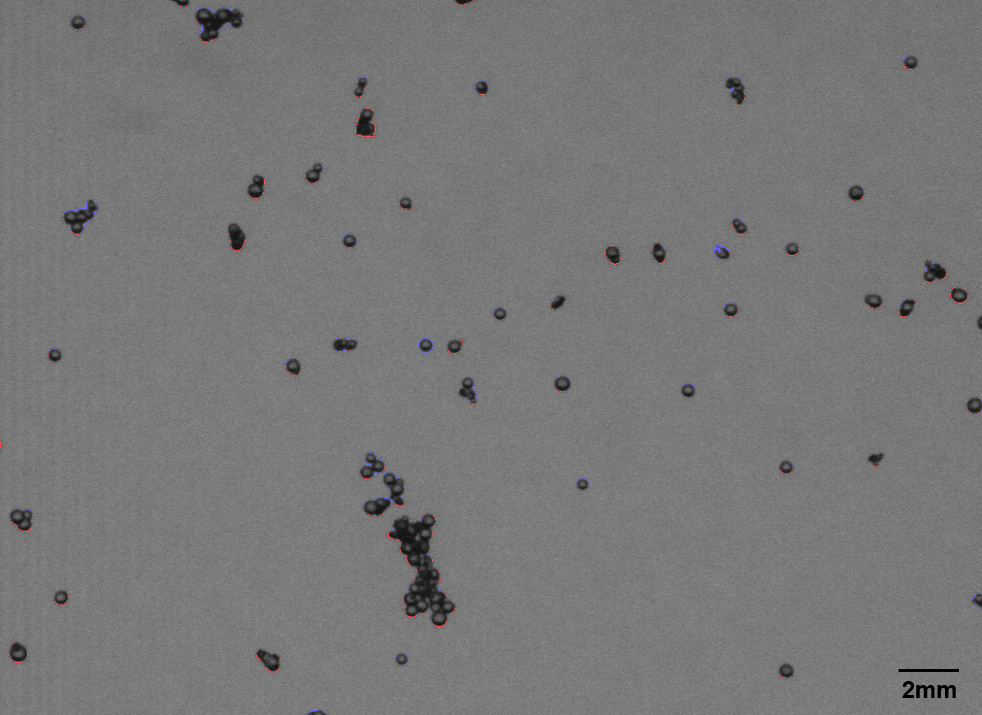}
    \caption{\textbf{Segmentation quality.} The difference between the reference ground truth segmentation $p^*$ obtained using SAM~\cite{kirillov2023segment} and the method described in Section~\ref{sec:segmentations} is shown. Pixels segmented by the proposed method but not by SAM are highlighted in red, while those segmented by SAM but not by the proposed method are shown in blue.}
    \label{fig:segmentation_difference}
\end{figure}

To evaluate the quality of the segmentation procedure described in Section~\ref{sec:segmentations}, a combination of visual and quantitative analyses was performed. To do so, for five of the Camsizer images, a ground truth phase-wise segmentation $p^*\colon \pixPos \to \{0,1,2\}$ was generated by using a much slower state-of-the-art segmentation modelfrom the field of machine learning~\cite{kirillov2023segment} . However, this method is not feasible for inline segmentation due to its computational complexity in memory and time. The difference of this ground truth segmentation $p^*$, and the segmentation achieved with the method described in Section~\ref{sec:segmentations} is visualized in Figure~\ref{fig:segmentation_difference}. It can be observed that all objects are detected correctly, and differences in segmentations are due to small variations of the objects' outlines.  

Furthermore, for a quantitative evaluation of the segmentation quality, the segmentations $p$ from Section~\ref{sec:segmentations} are compared with $p^*$ by means of the intersection over union (IoU) metric~\cite{jaccard1912distribution}.
The IoU is defined as
\begin{align}
    \mathrm{IoU}(p,p^*)=
    \frac{
        |\{x \in \pixPos \colon p(x)=1 \text{ and } p^*(x)=1\}|
    }{
        |\{x \in \pixPos \colon p(x)=1 \text{ or } p^*(x)=1\}|
    }.
\end{align}
The segmentation method described in Section~\ref{sec:segmentations} achieves an average IoU score of 0.93 compared to the reference ground truth segmentation, indicating a high degree of agreement.

\subsection{Classification}\label{sec:results_classification}

For object type classification, a random forest is trained as described in Section~\ref{sec: classification} on basis of 1854 descriptor vectors and corresponding hand labeled object classes. Note that this training data consists of descriptor vectors computed from image data derived from both Experiment A and Experiment E as well as all time steps in $T$. The grid-search to tune the hyperparameters leads to a random forest with $100$ decision trees each of which have a  maximal depth of $5$. Moreover, $|J|=5$ randomly chosen descriptors  are considered per tree.
The prediction quality of the random forest classifier that achieved the best results, is evaluated based on a second set of hand-labeled particle descriptors, again containing descriptors from all experiments and time steps, but not used in the training of the random forest. Specifically, the prediction quality of the classifier was evaluated based on descriptor vectors of 276 primary particles 133 chain-like agglomerates and 148 raspberry-like agglomerates. The corresponding confusion matrix is shown in Table~\ref{tab:confusion_matrix}~(left).

To assess the added value of utilizing a larger number of particle descriptors in the random forest classification compared to using only a few, a second, more interpretable classification approach based on only two particle descriptors was implemented. Specifically, the diameter $\diameter$ and eccentricity $\eccentricity$ were selected  for this classification, as the diameter is a straight-forward descriptor to distinguish between primary particles and agglomerates. Furthermore, chain-like agglomerates tend to be much more elongated than raspberry-like agglomerates, thus, typically showing higher values of $\eccentricity$.
Based on the data used for the training of the random forest classifier, an optimal threshold for the area-equivalent diameter $\diameter$ was computed to distinguish primary particles from agglomerates. Note that it is enough to restrict the search for such a threshold to the observed values of $\diameter$ in the training data, ensuring both efficiency and straightforward computation. Subsequently, a second optimal threshold for the eccentricity descriptor $\eccentricity$ was determined to classify non-primary particles into chain-like and raspberry-like agglomerates.  The resulting confusion matrix for this reference classifier, computed on the validation data, is shown in Table~\ref{tab:confusion_matrix}~(right). The random forest classification, operating on all 22 descriptors considered in the present paper, achieves a much better precision  among all classes than the reference classifier which only considers the diameter $\diameter$ and eccentricity $\eccentricity$. This justifies the use of a more complex, and thus, more time-consuming, classification procedure.

The influence of individual  descriptors on the classification of particles/agglomerates can be measured by so-called Shapley values~\cite{NIPS2017_8a20a862,lundberg2020local2global,Shap_Bee}. It turns out that for classifying primary particles, the most influential descriptor is roundness $\circularity$, followed by the length of the major axis $\axisMajorLength$ and the ratio of the radii of the inscribed and enclosing spheres $\radiiRatio$. For identifying chain-like agglomerates, roundness $\circularity$ remains the most influential descriptor, followed by the lengths of the minor axis $\axisMinorLength$ and major axis $\axisMajorLength$. Finally, for classifying raspberry-like agglomerates, the minor axis length $\axisMinorLength$ is the most influential descriptor, followed by roundness $\circularity$ and again the major axis length $\axisMajorLength$

\begin{table}[H]
\caption{\textbf{Confusion matrices.} The confusion matrix for the random forest classifier (left) and of the corresponding reference classifier that is based on the diameter $\diameter$ and eccentricity $\eccentricity$  is shown. The term GT refers to the particle classes assigned by hand labeling.}
    \label{tab:confusion_matrix}
\begin{subfigure}{0.48\textwidth}
    \centering
    \begin{tabular}{|c|c|c|c|c|}
    \cline{3-5}
     \multicolumn{2}{c|}{}& \multicolumn{3}{c|}{random forest} \\
    \cline{3-5}
    \multicolumn{2}{c|}{}&primary&chain&berry\\
    \hline
    
    \multirow{3}{*}{\rotatebox{90}{GT}}&
         primary& 275&0&0\\ 
   \cline{2-5}
         &chain&0&123&10\\
         \cline{2-5}
         &berry&0&11&137\\
         \hline
    \end{tabular}
\end{subfigure}
\begin{subfigure}{0.48\textwidth}
    \centering
    \begin{tabular}{|c|c|c|c|c|}
    \cline{3-5}
     \multicolumn{2}{c|}{}& \multicolumn{3}{c|}{reference classifier} \\
    \cline{3-5}
    \multicolumn{2}{c|}{}&primary&chain&berry\\
    \hline
    
    \multirow{3}{*}{\rotatebox{90}{GT}}&
         primary& 265&0&11\\ 
   \cline{2-5}
         &chain&13&96&24\\
         \cline{2-5}
         &berry&0&37&111\\
         \hline
    \end{tabular}
\end{subfigure}
    
\end{table}

\subsection{Particle descriptor prediction}
\label{sec: result_descripotr}

\subsubsection{Temporal evolution of class sizes}\label{sec:results_regression_size}
\begin{figure}[H]
\centering
\begin{subfigure}{0.4\textwidth}
    \includegraphics[width=\textwidth]{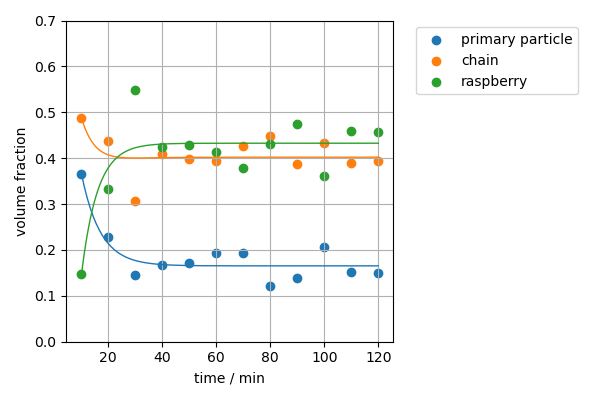}
\end{subfigure}
%\hfill
\begin{subfigure}{0.4\textwidth}
    \includegraphics[width=\textwidth]{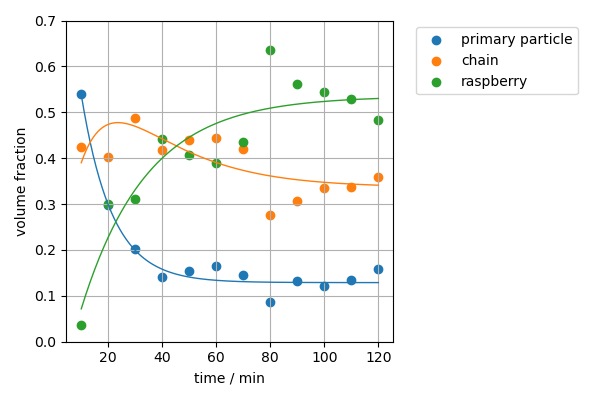}
\end{subfigure}
       
\caption{\textbf{Temporal evolution of classes sizes.} The temporal evolution of the area-weighted size fraction for different particle/agglomerate classes in Experiment A (left) and Experiment E (right) is shown. Circles represent the area-weighted fraction of particles/agglomerates observed in the image data. The fitted regression function (Equation~\eqref{eq:regressionfunction}), derived in Section~\ref{sec:results_regression_size}, is shown as curves in the corresponding colors.}
\label{fig:vol_frac}
\end{figure}

In the following the agglomeration process in both Experiments A and E is modeled in terms of the size fractions of the three object classes over time. More specifically, for each experiment and each measured time step, the fraction of segmented foreground pixels belonging to the respective classes is first computed. Then, as described in Section~\ref{sec: class_sizes}, the regression function from Equation~\eqref{eq:regressionfunction} is fitted to model the evolution of class size fractions over time for each experiment individually. The resulting fitted regression functions are shown in Figure~\ref{fig:vol_frac}. 

Focusing on primary particles in Experiments A and E, their number is decreasing as more and more primary particles being agglomerated, see Figure~\ref{fig:vol_frac} (blue lines). Nevertheless, the fraction of primary particles does not vanish completely, due to two effects: 1) continuous feed of primary particles into the process, 2) intermediate breakage of formed agglomerates.
Volume fractions of chain-like and raspberry-like agglomerates also attain steady values, although with different dynamics. Initially, a large number of chain-like agglomerates are formed (only few agglomeration events required) which are then integrated into larger, raspberry-like structures. In total, a slight prevalence of raspberry-like agglomerates over chain-like agglomerates is observed in Experiment A, that can be attributed to the higher mechanical stability of raspberry-like agglomerates (larger number of solid bridges with surrounding particles). Although the general trends are the same for Experiment E, it differs in its kinetics, i.e., agglomerate formation is slower in Experiment E than in Experiment A. This can be related to the operation conditions: Experiment E  is operated at a higher gas inlet temperature and with a higher binder content. Consequently, the sprayed droplets will dry faster and form individual particles (called overspray) that do not contribute to the agglomeration, as these pre-dried droplets will not deposit on the primary particles or agglomerates.

\subsubsection{Fitted univariate distributions over time}
\label{sec: res_univ}

To analyze and model not only the volume fraction of these classes, but also the  probability distributions of size and shape descriptors, Section~\ref{sec: parametric_modeling} introduced a parametric modeling procedure for the univariate distributions of object size $\diameter$ and solidity $\solidity$. We apply the described procedure to the data of each class and of Experiment A and E, to obtain the univariate fits $f^{\widehat{G}_\diameter,\widehat{\omega}_{t,d}}_{d}$  and$f^{\widehat{G}_s,\widehat{\omega}_{t,s}}_{s}$ for each $t \in T$. The resulting best-fitting family and the corresponding parameters of the marginal distributions for each experiment, class and time step are presented in Table~\ref{tab:margins}.

For the primary particles, no trend can be seen in the evolution of distribution parameters over time for both experiments. This indicates that the size and shape of these particles remain unchanged over time, which is expected for primary particles. Consequently, fitting a regression line provides no additional value as we assume that the univariate distributions are constant over time.

\begin{figure}[H]
    \centering
    \includegraphics[width=0.9\linewidth]{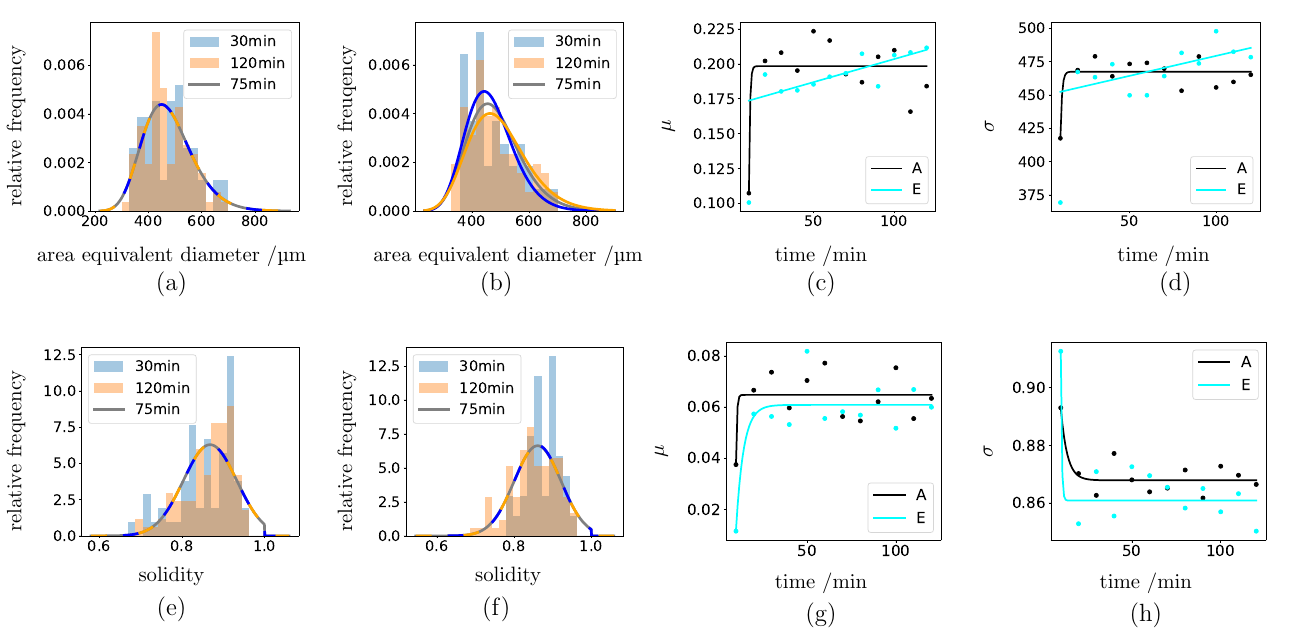}
    \caption{\textbf{Temporal evolution of marginal distributions.} Parametric modeling of marginal distributions. The marginal distributions of the area-equivalent diameter $\diameter$ (upper row) and solidity $\solidity$ (lower row) for both, Experiment A (column 1) and Experiment E  (column 2) are shown for the time steps 30 min, 75 min and 120 min. The distributions of all time steps are obtained by parameter regression. The parameters of the marginal distributions and a fitted regression line are shown (column 3-4) for both experiments. The regression makes it possible to make statements about unobserved time steps, e.g. for 75 minutes for which no data is available.}
    \label{fig: marginal_dist_berry}
\end{figure}

The time-dependent regression function given in  Equation \eqref{eq:regressionfunction} is fitted to the distribution parameters given in Table~\ref{tab:margins}, i.e., to  distribution parameters of each experiment, each agglomerate type and each descriptor, by minimizing Equation~\eqref{eq: loss_paramter_fitting}. The resulting regression functions fitted to the distribution parameters in Table~\ref{tab:margins} for raspberry-like agglomerates are shown on the right side of Figure~\ref{fig: marginal_dist_berry}. The upper row shows the results corresponding to the area-equivalent diameter $\diameter$, whereas the lower row shows the results corresponding to the solidity $\solidity$. All parameters reach saturation after 30 minutes, with the exception of the parameters that describe the area-equivalent diameter of raspberry-like agglomerates in Experiment E. The fitted regression functions can be used to predict marginal distributions for unmeasured time steps. 

Figures~\ref{fig: marginal_dist_berry} a), b), e) and f) visualize the resulting marginal distributions of raspberry-like agglomerates for experiments A and E at the time steps 30 min and 120 min. Note that the parameters of the visualized probability densities (lines) have been obtained by using the prediction of the associated  regression functions at these time steps.  As can be observed, the distributions correspond well with the available data.  Moreover, the regression functions have been used to predict distribution parameters for the probability distribution of $d$ and $s$ at the time step 75 min, i.e., for a time step for which no data is available. The corresponding probability densities are visualized in gray in Figure~\ref{fig: marginal_dist_berry}. Interestingly, the distributions obtained from the regression curves remain unchanged between these time steps, except for the area-equivalent diameter $\diameter$ of agglomerates in Experiment E. However, in Experiment E, the size of the agglomerates continues to increase after 30 min, while their shape, i.e., their solidity $\solidity$, remains constant. This behavior is to be expected based on the regression functions for the distribution parameters.

\begin{table}[ht]
\footnotesize
    \centering
    \caption{\textbf{Fitted parametric univariate distributions.} Parametric families of univariate distributions that have been identified as suitable for modeling the univariate probability densities of $d$ and $s$ for each class in Experiments A and E  across all time steps, along with the fitted parameters for each time step (desc. = descriptor, dist. = distribution, para. = parameter).}
    \label{tab:margins}
    \begin{tabular}{cccccccccccccccc}
        \toprule
          \multirow{2}{*}{desc.} & \multirow{2}{*}{dist.} & \multirow{2}{*}{par.} & \multicolumn{11}{c}{time step $t$ /min} \\
           & & & 10 & 20 & 30 & 40 &50 & 60 &70&80& 90 & 100 & 110 & 120 \\
           \midrule \multicolumn{14}{c}{Experiment A; primary particle} \\ \midrule 
  \multirow{2}{*}{$\diameter$} & \multirow{2}{*}{normal}  & $ \mu=$  & 215.3 & 224.89 & 222.79 & 221.18 & 221.24 & 221.3 & 222.46 & 223.22 & 226.97 & 223.83 & 226.57 & 221.81\\ 
 &  & $\sigma= $  & 29.01 & 28.53 & 29.6 & 29.58 & 27.9 & 27.77 & 27.82 & 27.24 & 29.6 & 30.34 & 29.08 & 29.05 \\  \multirow{2}{*}{$\solidity$} & \multirow{2}{*}{normal} &  $\mu =$  & 0.96 & 0.96 & 0.96 & 0.96 & 0.96 & 0.96 & 0.96 & 0.96 & 0.96 & 0.96 & 0.96 & 0.96\\ 
& & $\sigma = $  & 0.01 & 0.01 & 0.01 & 0.01 & 0.01 & 0.01 & 0.01 & 0.01 & 0.01 & 0.01 & 0.01 & 0.01\\ 
\midrule \multicolumn{14}{c}{Experiment A; chain} \\ \midrule 
  \multirow{2}{*}{$\diameter$} & \multirow{2}{*}{gamma}  & $ \alpha=$  & 33.53 & 33.57 & 42.44 & 24.78 & 25.99 & 25.63 & 26.08 & 21.5 & 30.3 & 25.01 & 32.85 & 25.83\\ 
 &  & $\beta= $  & 8.45 & 8.9 & 7.36 & 12.2 & 11.73 & 11.79 & 11.39 & 14.84 & 10.3 & 12.43 & 9.09 & 11.88 \\  \multirow{2}{*}{$\solidity$} & \multirow{2}{*}{normal} &  $\mu =$  & 0.91 & 0.91 & 0.91 & 0.9 & 0.9 & 0.9 & 0.91 & 0.89 & 0.9 & 0.9 & 0.91 & 0.91\\ 
& & $\sigma = $  & 0.04 & 0.04 & 0.06 & 0.05 & 0.05 & 0.06 & 0.06 & 0.07 & 0.05 & 0.06 & 0.06 & 0.06\\ 
\midrule \multicolumn{14}{c}{Experiment A; raspberry} \\ \midrule 
  \multirow{2}{*}{$\diameter$} & \multirow{2}{*}{log-normal}  & $ \mu=$  & 0.11 & 0.2 & 0.21 & 0.2 & 0.22 & 0.22 & 0.19 & 0.19 & 0.21 & 0.21 & 0.17 & 0.18\\ 
 &  & $\sigma= $  & 417.45 & 468.71 & 479.03 & 464.03 & 473.36 & 474.12 & 469.77 & 453.2 & 478.91 & 455.65 & 459.8 & 465.19 \\  \multirow{2}{*}{$\solidity$} & \multirow{2}{*}{normal} &  $\mu =$  & 0.89 & 0.87 & 0.86 & 0.88 & 0.87 & 0.86 & 0.87 & 0.87 & 0.86 & 0.87 & 0.87 & 0.87\\ 
& & $\sigma = $  & 0.04 & 0.07 & 0.07 & 0.06 & 0.07 & 0.08 & 0.06 & 0.05 & 0.06 & 0.08 & 0.06 & 0.06\\ 
\midrule \multicolumn{14}{c}{Experiment E; primary particle} \\ \midrule 
  \multirow{2}{*}{$\diameter$} & \multirow{2}{*}{normal}  & $ \mu=$  & 208.44 & 213.41 & 217.9 & 219.66 & 217.17 & 219.35 & 216.03 & 217.14 & 217.75 & 217.13 & 215.05 & 213.89\\ 
 &  & $\sigma= $  & 31.73 & 28.96 & 28.5 & 29.66 & 31.07 & 27.61 & 29.96 & 27.58 & 27.51 & 31.68 & 31.77 & 30.52 \\  \multirow{2}{*}{$\solidity$} & \multirow{2}{*}{normal} &  $\mu =$  & 0.96 & 0.96 & 0.96 & 0.96 & 0.96 & 0.96 & 0.96 & 0.96 & 0.96 & 0.96 & 0.96 & 0.96\\ 
& & $\sigma = $  & 0.02 & 0.01 & 0.01 & 0.01 & 0.01 & 0.01 & 0.01 & 0.01 & 0.01 & 0.01 & 0.02 & 0.02\\ 
\midrule \multicolumn{14}{c}{Experiment E; chain} \\ \midrule 
  \multirow{2}{*}{$\diameter$} & \multirow{2}{*}{log-normal}  & $ \mu=$  & 0.18 & 0.2 & 0.19 & 0.2 & 0.2 & 0.21 & 0.21 & 0.18 & 0.21 & 0.22 & 0.19 & 0.23\\ 
 &  & $\sigma= $  & 256.34 & 280.19 & 293.77 & 303.7 & 304.58 & 310.27 & 300.08 & 306.56 & 292.69 & 307.77 & 305.23 & 295.43 \\  \multirow{2}{*}{$\solidity$} & \multirow{2}{*}{normal} &  $\mu =$  & 0.91 & 0.91 & 0.9 & 0.9 & 0.9 & 0.89 & 0.9 & 0.9 & 0.9 & 0.89 & 0.89 & 0.9\\ 
& & $\sigma = $  & 0.04 & 0.06 & 0.06 & 0.06 & 0.06 & 0.06 & 0.06 & 0.06 & 0.06 & 0.06 & 0.05 & 0.06\\ 
\midrule \multicolumn{14}{c}{Experiment E; raspberry} \\ \midrule 
  \multirow{2}{*}{$\diameter$} & \multirow{2}{*}{log-normal}  & $ \mu=$  & 0.1 & 0.19 & 0.18 & 0.18 & 0.19 & 0.19 & 0.19 & 0.21 & 0.18 & 0.21 & 0.21 & 0.21\\ 
 &  & $\sigma= $  & 369.2 & 467.29 & 463.34 & 473.1 & 449.76 & 449.77 & 464.17 & 481.65 & 473.71 & 497.88 & 482.48 & 478.38 \\  \multirow{2}{*}{$\solidity$} & \multirow{2}{*}{normal} &  $\mu =$  & 0.91 & 0.85 & 0.87 & 0.86 & 0.87 & 0.87 & 0.87 & 0.86 & 0.87 & 0.86 & 0.86 & 0.85\\ 
& & $\sigma = $  & 0.01 & 0.06 & 0.06 & 0.05 & 0.08 & 0.06 & 0.06 & 0.06 & 0.07 & 0.05 & 0.07 & 0.06\\ 
\bottomrule
\end{tabular}
    
\end{table}

We obtain similar results for chain-like agglomerates. For the solidity $\solidity$, the distribution parameters remain almost unchanged after 30 min for both experiments, see Table~\ref{tab:margins}. The same applies for the area-equivalent diameter $\diameter$ and Experiment A. However, similar to the raspberry-like agglomerates, the area-equivalent diameter in Experiment E continues to increase beyond 30 minutes, as indicated by the rising value of $\sigma$ in Table~\ref{tab:margins}. Nevertheless, this increase is less pronounced than in the raspberry-like agglomerates and nearly reaches saturation after 60 minutes.

The general preservation of shape (solidity) with respect to the operation conditions is not surprising: Due to fluidization, agglomerates and recently attached primary particles undergo constant collisions with other agglomerates and the apparatus walls. The mechanical stress is sufficient to overcome the cohesive forces required to break single bridges. This favors agglomerate structures with low surface areas and large numbers of contact points between primary particles, i.e., the observed raspberry-like shape.
Although the shape is mostly unchanged, the dynamics of agglomerate formation, agglomerate size and the intermediate volume fraction differ significantly, compare Figure~\ref{fig:vol_frac}.

\subsubsection{Fitted bivariate distributions over time}

So far, only univariate distributions of $\diameter$ and $\solidity$ have been considered, but these do not model the dependency of the two descriptors. Thus, we want first to investigate the dependency of area-equivalent dimeter and solidity for all three classes observed in Experiments A and E. A non-parametric measure  for dependence is empirical Kendall's tau, see~\cite{joe2014dependence}, which takes values in the interval  $[-1,1]$, where a value close to zero indicates that there is no clear positive or negative correlation between the area-equivalent diameter and solidity.

The empirical Kendall's tau values between $d$ and $s$ for primary particles, averaged over all time steps,  is 0.10 across both experiments. For chain-like agglomerates, the corresponding average value of empirical Kendall's tau over all time steps are  $-0.35$ and $-0.38$ for Experiments A and E, respectively. Analogously, we determined the averages of $-0.47$ and $-0.40$ for raspberry-like agglomerates for Experiments A and E over all time steps, respectively. Due to the average Kendall's tau being close to 0, we assume independence of $\diameter$ and $\solidity$ for primary particles. 

Recall that copulas can be used to model the dependencies between two descriptors. Section~\ref{sec: parametric_modeling} outlined how the marginal distributions and an Archimedean copula can be used to model the bivariate distribution of $d$ and $s$. However, since we assume independence for the two descriptors in the case of primary particles, we solely employ copulas to model the bivariate distributions of $d$ and $s$ for agglomerates. 

To determine the best-fitting copula family and the corresponding parameter for each time step of the agglomerates, we apply the methods from Section~\ref{sec: parametric_modeling} to the data of chain-like and raspberry-like agglomerates for Experiment A and Experiment E to obtain a sequence of parametric copulas $c_{\mathrm{\diameter, \solidity}}^{\widehat{Z}, \widehat{\theta}_t}$, see Table~\ref{tab:copula_parameters}.

\begin{table}[H]
\small
\caption{\textbf{Fitted parametric copula functions.} Archimedean copulas that have been identified as suitable for modeling the bivariate probability of $\diameter$ and $\solidity$ for raspberry-like and chain -like agglomerates in Experiment A and Experiment E across all time steps, along with the fitted parameters for each time step (rot. = rotation, exp. =experiment, par. = parameter, ali. = Ali-Mikhail-Haq).}
    \label{tab:copula_parameters}
    \centering
    \begin{tabular}{p{0.8cm}cccccccccccccccc}
        \toprule
           \multirow{2}{*}{exp.} & \multirow{2}{*}{type} & \multirow{2}{*}{copula}  & \multirow{2}{*}{rot.} &  \multirow{2}{*}{par.} & \multicolumn{11}{c}{time step $t$ /min} \\
           & & & & &  10 & 20 & 30 & 40 &50 & 60 &70&80& 90 & 100 & 110 & 120 \\ \midrule
      A& chain& ali&90& $\theta = $ & 0.87 & 0.87 & 0.98 & 1.0 & 0.98 & 0.99 & 0.99 & 0.99 & 1.0 & 1.0 & 0.99 & 0.96\\ 
A& raspberry&clayton&270& $\theta = $ & 0.78 & 1.88 & 1.72 & 1.55 & 1.83 & 1.79 & 1.53 & 1.97 & 2.2 & 1.92 & 1.53 & 1.8\\ 
E& chain& ali&90& $\theta = $ & 0.79 & 0.96 & 0.98 & 1.0 & 0.99 & 0.99 & 0.98 & 0.97 & 0.96 & 0.99 & 0.96 & 0.98\\ 
E& raspberry& clayton&270& $\theta = $ & 0.16 & 1.22 & 1.5 & 1.4 & 1.77 & 1.81 & 1.6 & 1.17 & 1.58 & 1.55 & 1.84 & 1.81\\ 
\bottomrule
\end{tabular}
    
\end{table}

\begin{figure}[H]
    \centering
    \includegraphics[width=0.9\linewidth]{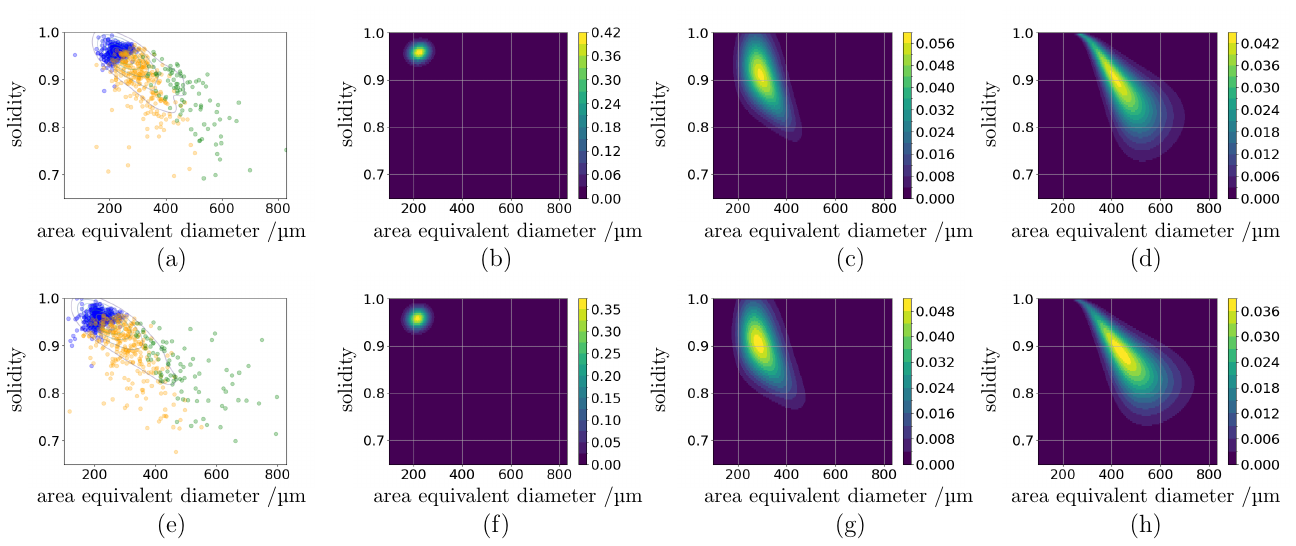}
    \caption{\textbf{Bivariate distribution of descriptors.} Bivariate distribution of area-equivalent diameter and solidity for particles/agglomerates observed in Experiments A (upper row) and E (lower row). The bivariate distributions of all measured particles is represented using iso-lines of the probability density obtained by means of a kernel density estimation (column 1) at time step 120 min. Furthermore, a decomposition of this distribution into the parametric distributions of the classes of primary particles, chain-like agglomerates, and raspberry-like agglomerates is shown (columns 2-4).}
    \label{fig: biv_classes}
\end{figure}

Figure~\ref{fig: biv_classes} visualizes 
exemplarily chosen bivariate probability densities of $d$ and $s$ for primary particles, chain-like agglomerates and raspberry-like agglomerates. A visual inspection of the densities for each class shows that a linear combination of the three densities represents the distribution of all scatter points shown in Figure~\ref{fig: biv_classes} (first column) well.

\begin{figure}[H]
    \centering
    \includegraphics[width=0.9\linewidth]{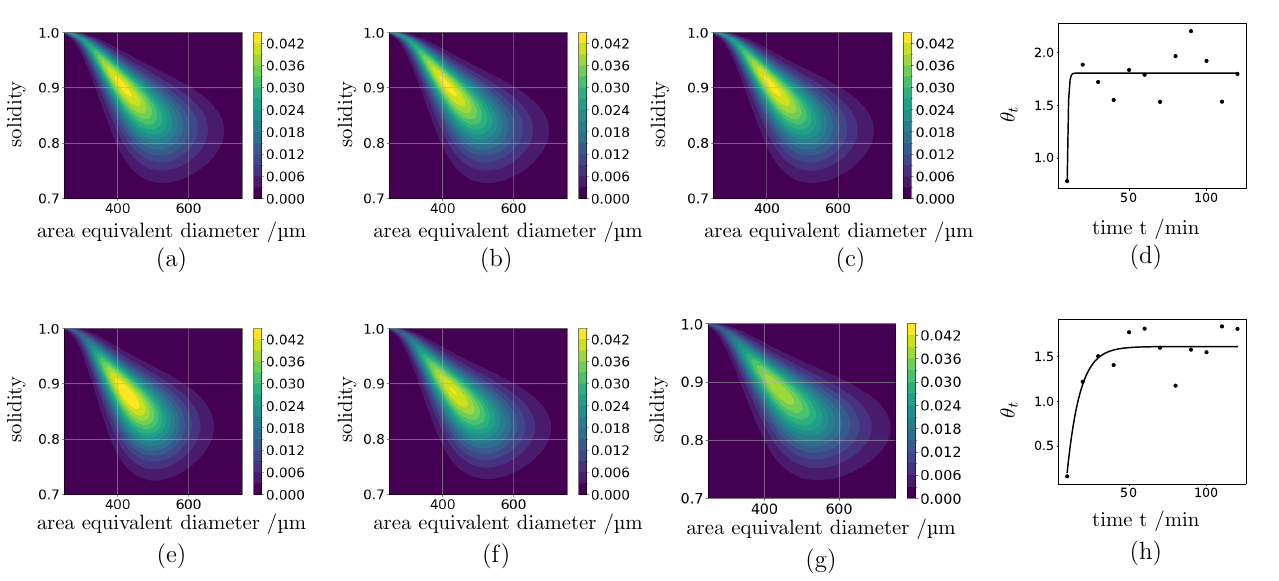}
    \caption{\textbf{Temporal evolution of marginal distributions.} Evolution of bivariate probability density of area-equivalent diameter and solidity in time for Experiment A (upper row) and E (lower row). The bivariate probability density for raspberry-like agglomerates is shown for the  time steps of 30 min (column 1) and 120 min (column 3). Furthermore, the predicted bivariate probability density for raspberry-like agglomerates is shown for the time step of 75 min (column 2) for which no measured data is available.}
    \label{fig: biv_regression}
\end{figure}

As for the distribution parameters of the marginal distributions of $\diameter$ and $\solidity$, temporal changes of copula parameters can be analyzed for agglomerates. Therefore, time-dependent regression functions as given in Equation~\eqref{eq:regressionfunction} are fitted to the values of distribution parameter  by minimizing Equation~\eqref{eq: loss_paramter_fitting}.  
The resulting regression functions fitted to the copula parameters in Table~\ref{tab:copula_parameters} for raspberry-like agglomerates observed in Experiments A and E are shown in Figures~\ref{fig: biv_regression}d) and h). The parameters of Experiment A reach saturation after 20 min, while the parameters of Experiment E   reach saturation after 40 min. In addition, the bivariate probability densities of $d$ and $s$ for raspberry-like agglomerates after 30 and 120 min is shown in the first and third column,
where the copula parameters of the distributions have been determined by means of regression. Once again, no significant changes can be observed in the distributions for Experiment A. However, in Experiment E, an even more pronounced shift in the distributions can be seen, which is mainly caused by the parameter evolution of the area-equivalent diameter, see Section~\ref{sec: res_univ}.

Similarly to the univariate case, the fitted regression functions can be used to estimate copula parameters and consequently bivariate distributions of descriptor vectors for unmeasured time steps.
The predicted bivariate distributions for raspberry-like agglomerates are shown in Figures~\ref{fig: biv_regression}b) and f) for Experiment A and Experiment E  at 75 min, respectively.

For chain-like agglomerates, the parameters in both experiments reach saturation after 30 min. Results therefore indicate that with respect to structure formation, dynamic equilibrium between agglomeration of primary particles into chain-like agglomerates, transition from chains-like agglomerates to raspberry-like agglomerates and the breakage of too large raspberry-like agglomerates is reached. This timescale is also in line with results on average agglomerate size presented in~\cite{Aisel2024}. 
The previously presented results on modeling the evolution of particle and agglomerate structure distributions over time provide valuable insights into the process.

\subsection{Sensitivity of fitting procedure of copula-based bivariate distributions}\label{sec: results_Sensitivity}

As the previous results were obtained using a large amount of data that is not available in an online or real-time measurement situation, the question arises how sensitive these results are with respect to the number of evaluated objects. Or: How many objects must be evaluated before a reliable and robust estimate of the multi-dimensional structure of agglomerates is obtained? 

In order to investigate the sensitivity of the model fitting procedure described in Section~\ref{sec: result_descripotr}, we deployed the sensitivity analysis procedure described in Section~\ref{sec:sensitivity} to the particle descriptors measured in Experiment E at time 120 min. 
More precisely, for each object class we conducted the sensitivity analysis described in Section~\ref{sec:sensitivity} by setting $\mathcal{Y}$ as the set of all descriptor vectors $(\diameter,\solidity)\in [0,\infty) \times [0,1]$ of particles/agglomerates of the respective class, measured in Experiment E at time 120 min. Then, for each $\nBootstrap\in \{5,20,35,\ldots,140\}$ a total of 1\,000 bootstrap samples $\widetilde{\mathcal{Y}}$  of size $\nBootstrap$ are independently generated from $\mathcal{Y}$ and the corresponding models $\bootstrapDensity$ are fitted. Note that since we assume that the parametric families for the marginal distributions and for the copula do not change for different time steps within one experiment and object class, the parametric families associated with $\bootstrapDensity$ are chosen in accordance with 
Table~\ref{tab: univariate_dist} and Table~\ref{tab:copula_parameters}.
More precisely, when fitting $\bootstrapDensity$ to $\widetilde{\mathcal{Y}}$, we skip the search for optimal parametric families for both the marginal distributions and the copula, as we take the families from $f_\mathrm{d,s}$ given in Table~\ref{tab: univariate_dist} and Table~\ref{tab:copula_parameters}. We then optimize only their parameters, $\omega_d,\omega_s\in\R^2$ and $\theta\in \Theta$ with respect to $\widetilde{\mathcal{Y}}$, using maximum likelihood estimation~\cite{aitkin2010statistical}.

\begin{figure}[H]
    \centering
    \includegraphics[width=0.9\linewidth]{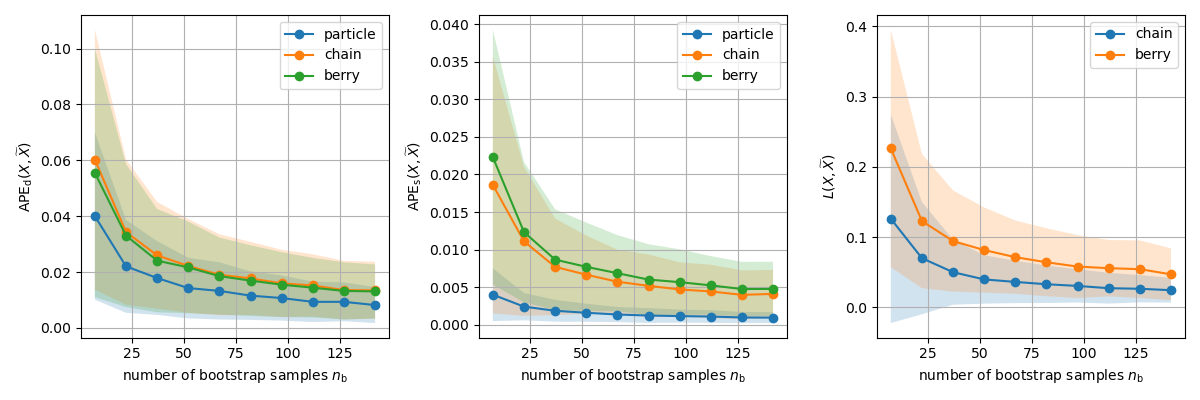}
    \caption{\textbf{Bootstrap sampling-based sensitivity.} For all particle types, the similarity of $\density$ computed by means of all data descriptor vectors measured in Experiment E  at time step 120 min and $\bootstrapDensity$ computed by means of a subset of $\nBootstrap$ descriptor vectors is shown. On the left and in the middle, the absolute percentage loss (see Equation~\eqref{eq:apeds} and Equation~\eqref{eq:apeds}) is shown for different sizes $\nBootstrap$. On the right-hand side, the similarity of  $\density$ and  $\bootstrapDensity$ is shown by means of the integral over the absolute difference of the respective copula functions, see Equation~\eqref{eq:copula_loss}. All experiments are conducted, 1\,000 times, the solid lines show the arithmetic mean, while the shaded area shows the standard deviation.}
    \label{fig:bootstrap}
\end{figure}

The results of this analysis are presented in Figure~\ref{fig:bootstrap}. On the left side and in the middle, the similarity of $\density$ and $\bootstrapDensity$ are shown by means of the APE$_d$ and APE$_s$ introduced in Equation~\eqref{eq:apeds} and Equation~\eqref{eq:apes}. It can be observed that accurate modeling of primary particles requires only a few observations. In contrast, modeling chain-like agglomerates necessitates a larger number of observations, while the modeling of raspberry-like agglomerates requires the most observations. This aligns with the observation that raspberry-like structures are much more complex than primary particles or chain-like agglomerates.
On the right side of Figure~\ref{fig:bootstrap}, we show the sensitivity of the procedure to fit the dependency structure of the probability density in dependence of the amount of available data. Since the dependency structure of $d$ and $s$ for primary particles is assumed to be independent, they are excluded from this analysis. As before, the dependency structure of $d$ and $s$ for raspberry-like agglomerates is more sensitive than that of chain-like agglomerates.

As raspberry-like agglomerates have the most complex structure and appear less frequently in our data, estimates on the required number of identified objects are expressed with respect to this agglomerate class. Given a desired expected error, in terms of APE$_d$ and APE$_s$, the minimum number of detected agglomerates can be predicted. Figure~\ref{fig:bootstrap} (Experiment E) shows that approximately 70 raspberry-like agglomerates need to be measured in order to achieve in expectation an APE$_d$ and APE$_s$ of less than 2~\%. 

For Experiment E  with an average count of about 3.5 raspberry-like agglomerates per image, this corresponds to about 20 images. Given typical acquisition rates (frame rates) of about 60 images per second, this means that after approximately a third of a second of measurement time the quantity of information necessary for such a precision can be acquired.

These results are exceptionally important for implementing online measurement of agglomerate structure formation and online model-learning and process adaptation, e.g., closed-loop (feedback) control of fluidized bed spray agglomeration processes for defined agglomerate structures. In particular, it enables almost immediate action to steer synthesized agglomerate structures into preferable directions; furthermore, deviations and disturbances can be quickly identified and severe malfunctions (e.g., break-down of fluidization due to excessive agglomeration) can be prevented.

\section{Conclusion}
\label{sec: conclusion}

In this paper, the agglomeration process of glass beads in a SFB agglomeration over time is quantitatively investigated and modeled. In particular, the structure formation is investigated by analyzing inline images sequences followed by time-dependent multivariate stochastic modeling of size and shape descriptors of particles/agglomerates.

Particles and agglomerates are automatically segmented from image data, from which various geometrical and textural descriptors are computed. These descriptors are used to classify segmented objects as primary particles, chain-like agglomerates and raspberry-like agglomerates. For this task, a fast and robust random forest classifier is successfully applied. Subsequently, parametric bivariate distributions are fitted for each class and time step. This allows the regression of  model parameters, which enables us to make statements about the distribution of descriptor vectors at unobserved time steps and predict the time-dependent multivariate descriptor distribution of these particle classes. In addition, the volume fraction of each class over time is investigated. The results show that no further agglomeration can be observed after 30 minutes with a small amount of binder. With a higher amount of binder, however, further agglomeration is observed after 30 minutes. 

In addition, a sensitivity analysis is performed to quantify the amount of data required in order to adequately model the bivariate distributions that characterize the state of agglomeration. It is shown that a higher amount of data is required for raspberry-like agglomerates due to their more complex shape and less-frequent occurrence than chain-like agglomerates and primary particles. The results of the sensitivity analysis show the minimum data required for online tracking of agglomerate formation dynamics. This opens the door to online monitoring of structure formation and feedback control of SFB agglomeration with respect to disturbance identification and rejection.  As a result, stable operation can be maintained while ensuring predefined product properties such as the re-hydration capacity and kinetics of the agglomerated material.

\medskip
\textbf{Supporting Information} 
Supporting Information is available from the authors.

% Acknowledgements
\medskip
\textbf{Acknowledgements}\\
Funding of this work by Deutsche Forschungsgemeinschaft (DFG) (project IDs 504524147 and 504580586) within the priority programme PP 2364 Autonomous Particle Processes is gratefully acknowledged.

\bibliography{ref}{}
\bibliographystyle{elsarticle-num}

\end{document}